\documentclass[aps,prb,twocolumn,showpacs,superscriptaddress,amsmath,amssymb]{revtex4-1}

\usepackage{graphicx}
\usepackage{bm}
\renewcommand{\vec}[1]{\bm{#1}}

\begin{document}



\title{Frustrated spin chains in strong magnetic field: dilute two-component Bose gas regime }



\author{A. K. Kolezhuk}
\affiliation{Institute of Magnetism, National Academy of Sciences and
  Ministry of Education, 36-b Vernadskii av., 03142 Kiev, Ukraine}
\affiliation{Institute of High Technologies, T. Shevchenko Kiev
  National University, 64 Volodymyrska str., 01601 Kiev, Ukraine}

\author{F. Heidrich-Meisner} 
\affiliation{Department of Physics and   Arnold Sommerfeld Center for Theoretical Physics,
  Ludwig-Maximilians-Universit\"at M\"unchen, D-80333 M\"unchen, Germany}

\author{S. Greschner}
\author{T. Vekua}
\affiliation{Institut f\"ur Theoretische Physik, Leibniz Universit\"at
  Hannover, Appelstr. 2, 30167 Hannover, Germany}

\date{\today}

\begin{abstract}
We study the ground state of frustrated spin-$S$ chains in a strong
magnetic field in the immediate vicinity of saturation.  In strongly
frustrated chains, the magnon dispersion has two degenerate minima at
inequivalent momenta $\pm Q$, and just below the
saturation field the system can be effectively represented as a dilute
one-dimensional lattice gas of two species of bosons that correspond
to magnons with momenta around $\pm Q$.  We present a theory of 
effective interactions in such a dilute magnon gas that allows us to make quantitative
predictions for arbitrary values of the spin.  With the help of this
method, we are able to establish the magnetic phase diagram of
frustrated chains close to saturation and study phase transitions
between several nontrivial states, including a two-component Luttinger
liquid, a vector chiral phase, and phases with bound magnons.  We
study those phase transitions numerically and find a good agreement
with our analytical predictions.
\end{abstract}



\maketitle

\section{Introduction}
\label{sec:intro}

Frustrated spin systems have been an object of avid interest to
researchers for several decades. In low dimensions, the interplay of
frustration and quantum fluctuations proved to be favorable for
generating states with unconventional spin order, such as chiral, 
nematic, or general multipolar phases, as well as for the existence of various
disordered states.  A strong external magnetic field 
competes with the exchange interaction, inducing a rich variety of
phase transitions. An overview of this
rapidly progressing field may be found in Refs.\ \onlinecite{Diep2004book,Mila2011book}. 

A spin chain with competing nearest and next-nearest neighbor exchange
interaction represents a paradigmatic model of frustrated spin systems in
one dimension. Despite many years of extensive studies, it continues
to deliver fresh surprises.  The frustrated spin-$S$ chain model,
which will be the main subject of our study, is described by the
following Hamiltonian:
\begin{eqnarray} 
\label{hamS}
\mathcal{H}&=& \sum_{n} \Big\{ \frac{J_{1}}{2} (S_{n}^{+}S_{n+1}^{-}+S_{n}^{-}S_{n+1}^{+}) 
+ J_{1}^{z}S_{n}^{z}S_{n+1}^{z} -H S_{n}^{z}
\nonumber\\
& +& \frac{J_{2}}{2} (S_{n}^{+}S_{n+2}^{-} + S_{n}^{-}S_{n+2}^{+})
+J_{2}^{z}S_{n}^{z}S_{n+2}^{z} \Big\}, 
\end{eqnarray}
where $(S_{n}^{\pm},S_{n}^{z})$ are spin-$S$ operators acting at  site $n$
of a one-dimensional lattice,
 $J_{1}$, $J_{1}^{z}$ and $J_{2}$, $J_{2}^{z}$  are nearest-neighbor (NN) and
next-nearest neighbor (NNN) exchange interactions,
and $H$ is the external magnetic field.
In this paper, we are
interested in the frustrated case, so $J_{2}$ is chosen to be positive
 and the sign of
$J_{1}$ is arbitrary. 
It is convenient to use the quantity
\begin{equation} 
\label{beta} 
\beta=J_{1}/J_{2}
\end{equation}
as the frustration parameter. The system may be alternatively viewed
as two antiferromagnetic chains connected by  ferro- or
antiferromagnetic zigzag couplings $J_{1}$, $J_{1}^{z}$.

For $S=\frac{1}{2}$, the above model has been extensively studied. Its
isotropic ($J_{1}^{z}=J_{1}$, $J_{2}^{z}=J_{2}$) version has a rich
magnetic phase diagram exhibiting states with competing unconventional
orders, both for a ferromagnetic
\cite{chubukov91a,HM+06,Vekua+07,Hikihara+08,Sudan+09} and an antiferromagnetic
\cite{OkunishiTonegawa03,KV05,Mc+08,Okunishi08,Hikihara+10} NN
exchange interaction $J_{1}$.  In particular, the vector chirality
(VC), which is a quantum remnant of the classical helical spin order
and is equivalent to the local spin current, competes, on the one
hand, with multipolar orders characterizing quasi-condensates
consisting of multimagnon bound states, and, on the other hand, with
the two-component Tomonaga-Luttinger liquid (TLL2)\cite{Okunishi+99}
where the two components correspond to magnons with momenta around two
degenerate dispersion minima at inequivalent points in the Brillouin
zone.  The same set of phases has been found in the anisotropic
version of the model as well.\cite{HM+09}

The model \eqref{hamS} with $S=\frac{1}{2}$ and a ferromagnetic (FM)
NN coupling has been discussed recently in connection with the
description of several quasi one-dimensional magnetic materials such
as LiCuVO$_4$
(Ref.~\onlinecite{enderle05,Buettgen+07,Svistov+10,ZhitomirskyTsunetsugu10}),
Rb$_2$Cu$_2$Mo$_{3}$O$_{12}$ (Ref.~\onlinecite{hase04}),
Li$_2$ZrCuO$_4$ (Ref.~\onlinecite{drechsler07}), and anhydrous $\rm
CuCl_{2}$ (Ref.\ \onlinecite{banks+09}). 

For higher values of $S$, not much is known about the magnetic phase
diagram of frustrated chains. A theoretical approach based on
bosonization\cite{KV05} predicts that in the regime of two weakly
coupled antiferromagnetic (AF) chains ($|\beta| \ll 1$) a VC
phase completely fills the magnetic phase diagram for all nonvanishing
values of the
magnetization between zero and saturation. On the FM side
($\beta<0$), for spin-$S\ge 1$ chains right below the saturation
field, a recent large-$S$ analysis in the framework of a dilute
magnon gas approach\cite{Arlego+11} predicts the existence of either a
VC phase, for $\beta_{cr}(S)<\beta<0$, or a metamagnetic
magnetization jump, for $-4<\beta<\beta_{cr}(S)$, with
$\beta_{cr}(S\ge 6)=-4$.  Numerical studies of magnetization
curves\cite{HM+07} have established the universal presence of a
magnetization plateau at one third of the saturation value in
antiferromagnetic chains with $S=1$, $\frac{3}{2}$, and $2$, while a
metamagnetic jump at saturation has been found in FM chains
\cite{Arlego+11} for $S=1$ and $S=\frac{3}{2}$ (metamagnetic jumps
have been observed earlier in $S=\frac{1}{2}$ chains\cite{Sudan+09} as
well, particularly at saturation for $\beta \to -4$).  The numerical
analysis of correlation functions\cite{Mc+08,KMc09} has been limited
to cuts at the fixed value of the frustration parameter $\beta=1$, and
has led to the conclusion that the field-induced VC phase is present in AF
chains with higher $S$ as well (values up to $S=2$ have been
studied). The TLL2 phase has been observed in $S=\frac{1}{2}$
frustrated AF
chains\cite{Okunishi+99,OkunishiTonegawa03,Okunishi08,Hikihara+10}, in
anisotropic $S=\frac{1}{2}$ frustrated FM chain,\cite{HM+09} as well
as in $S=1$ chain with bilinear and biquadratic
exchange.\cite{FathLittlewood98,Okunishi+99} Indications of its
existence have also been found in the form of kinks in magnetization
curves in higher-$S$ frustrated AF chains.\cite{HM+07}

In this paper, we study the ground state of a strongly frustrated ($|\beta|<4$) spin-$S$ chain
(\ref{hamS}) in strong magnetic fields in the immediate vicinity of
saturation.  Just below the saturation field, the system can be
represented as a dilute lattice gas of magnons, and since in the case
of strong frustration the magnon dispersion has two degenerate minima
at inequivalent points $\pm Q$ in momentum space, this lattice gas is
effectively a two-component one.  We present a theory of  effective
interactions in such a dilute magnon gas that is especially adapted to
the one-dimensional case, does not involve an $1/S$ expansion, and thus
allows us to draw quantitative predictions for arbitrary spin.  We
establish the magnetic phase diagram of isotropic and anisotropic
frustrated  chains close to saturation and study phase transitions
between several nontrivial states, including TLL2 and VC phases, and
phases with bound magnons.  Particularly, we show that the TLL2 phase
appears in FM and AF chains for all values of the spin $S$.  We
complement our analytical results with density
matrix renormalization group\cite{white92,schollwoeck05} (DMRG)
simulations.  Specifically, we compute the central charge as a
function of $\beta$ (at fixed magnetization close to the saturation
value), which allows us to locate
phase transitions between one- and two-component states. In order to
identify those one-component states that have vector chiral order, we
further calculate the chirality. Our DMRG results for the transition
from the VC to the TLL2 phase are in a good agreement with our
theoretical predictions.

The structure of the paper is as follows: in
Sect.\ \ref{sec:efftheory} we  describe mapping of the spin problem to the dilute
two-component lattice Bose gas and present the theory allowing us to
calculate effective interactions in such a gas for general spin $S$. 
In Sect.\ \ref{sec:2magnon}, the two-magnon scattering problem  in
a frustrated chain is considered, and its links to the effective theory of
Sect.\ \ref{sec:efftheory} are established.
Section \ref{sec:chain-results} discusses the specific predictions of the theory
for frustrated spin chain models  with different $S$ and either isotropic or anisotropic exchange interactions,
while Sect.\ \ref{sec:numerics} presents the
results of numerical analysis and their comparison with analytical
predictions. Finally, Sect.\ \ref{sec:summary} contains a brief
summary.

\section{Effective  two-component Bose gas model}
\label{sec:efftheory}
We are interested in the interactions between magnons in the regime of
a dilute magnon gas, i.e., for values of the field $H$ just slightly below the
saturation field $H_{s}$. Because of the diluteness of the gas, two-body
interactions dominate.
It is convenient to use the Dyson-Maleev representation \cite{noteDM} for the
spin operators in  (\ref{hamS}):
\begin{eqnarray} 
\label{DysonMaleev} 
&& S_{n}^{+}=\sqrt{2S}b_{n},\quad S_{n}^{-}=\sqrt{2S}b_{n}^{\dag}\Big(1-\frac{b_{n}^{\dag}b_{n}^{\vphantom{\dag}}}{2S}\Big),\nonumber\\
&& S_{n}^{z}=S-b_{n}^{\dag}b_{n}^{\vphantom{\dag}},
\end{eqnarray}
where $b_{n}$ are bosonic operators acting at site $n$.
To enforce the constraint
$b_{n}^{\dag}b_{n}^{\vphantom{\dag}}\leq 2S$, one can add 
the infinite interaction term
to the
Hamiltonian, which reads:
\begin{equation} \label{infU} 
\mathcal{H}\mapsto
\mathcal{H}+U\sum_{n}:(b_{n}^{\dag}b_{n}^{\vphantom{\dag}})^{2S+1}:\;,\quad U\to+\infty,
\end{equation}
where $:(\ldots):$ denotes normal ordering. Obviously, at the
level of two-body interactions this term is important only for $S=\frac{1}{2}$.

Passing further to the momentum representation for bosonic operators, one can rewrite the
model (\ref{hamS}) in the following form:
\begin{equation}
\label{hamB} 
\mathcal{H}=\sum_{k}(\varepsilon_{k}-\mu)b^{\dag}_{k}b_{k}^{\vphantom{\dag}}
+\frac{1}{2L}\sum_{kk'q}V_{q}(k,k')b^{\dag}_{k+q}b^{\dag}_{k'-q}b_{k}^{\vphantom{\dag}} b_{k'}^{\vphantom{\dag}}.
\end{equation}
Here $L$ is the total number of lattice sites, the chemical potential
is $\mu=H_{s}-H$, where 
\[
H_{s}=2SJ_{2}(1+\beta^{2}/8)+2S(J_{1}^{z}+J_{2}^{z})
\]
is the saturation field\cite{commentSaturationField} 
for the case of a strong NNN interaction $|J_{1}|\leq 4J_{2}$, which is
of our main interest here.
The
magnon energy 
\begin{equation} 
\label{ek} 
\varepsilon_{k}=2S(J_{1}\cos k+J_{2}\cos 2k-J_{1}^{z}-J_{2}^{z})+H_{s}
\end{equation}
is defined in  a way that ensures that $\min \varepsilon_{k}=0$. 
The lattice constant has been set to unity. It is easy to convince oneself that
$\varepsilon_{k}$ does  not depend on $J_{1,2}^{z}$. 
The dispersion $\varepsilon_{k}$ has two
degenerate minima at $k=\pm Q$, where 
\[
Q=\arccos\big[-\beta/4\big].
\]
The two-body interaction $V_{q}(k,k')$ generally depends on
the transferred momentum $q$ as well as on the incoming momenta $k$, $k'$:
\begin{equation} 
\label{Vq} 
V_{q}(k,k')=\begin{cases} J^{z}(q) -\frac{1}{2}\big[ J(k)+J(k') \big]
& 
S\geq 1,\\
U+J^{z}(q),\quad U\to +\infty &  S=\frac{1}{2} 
\end{cases},
\end{equation}
where 
\begin{eqnarray} 
\label{Jk} 
J(k)&=&2J_{1}\cos k +2J_{2}\cos 2k,\nonumber\\
J^{z}(k)&=&2J_{1}^{z}\cos k +2J_{2}^{z}\cos 2k.
\end{eqnarray}
It is easy to see that in this notation
$\varepsilon_{k}=S\big[J(k)-J(Q) \big]$.

\begin{figure}[tb]
 \includegraphics[width=0.45\textwidth]{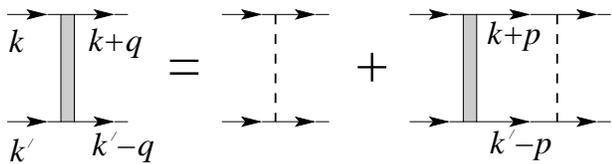}
 \caption{
\label{fig:Bethe-Salpeter} 
The ladder approximation to the Bethe-Salpeter equation for the renormalized two-body interaction vertex 
$\Gamma_{q}(k,k';E)$. Solid lines denote bare propagators. The approximation becomes exact at $\mu\to 0$.}
\end{figure}

The model (\ref{hamB}) defines a gas of bosons with a nontrivial
double-minima dispersion, on a one-dimensional (1d) lattice.
We are interested in the renormalized two-body interaction in this gas in the
dilute limit, i.e., $\mu\to 0$.
In the $\mu\to 0$ limit, the self-energy vanishes,
\cite{Uzunov81} so  the full propagator coincides with the bare one.
Then, the Bethe-Salpeter (BS) equation for the renormalized two-body interaction vertex 
$\Gamma_{q}(k,k';E)$ (where $E$ is the total energy of the incoming particles) in the limit $\mu\to 0$ takes the
following simple form:\cite{BatyevBraginskii84}
\begin{eqnarray} 
\label{BS-gen}
 \Gamma_{q}(k,k';E)&=&V_{q}(k,k')\\
&-&\frac{1}{L}\sum_{p}\frac{V_{q-p}(k+p,k'-p)\Gamma_{p}(k,k';E)}{\varepsilon_{k+p}+\varepsilon_{k'-p}-E}.\nonumber
\end{eqnarray}
This expression is schematically shown in terms of Feynman diagrams in Fig.~(\ref{fig:Bethe-Salpeter}).
For the
physical ``on-shell''  vertex the energy is fixed at
$E=\varepsilon_{k}+\varepsilon_{k'}$, but we keep 
Eq.~(\ref{BS-gen}) in a general form for reasons that will become 
 clear shortly. 

At a low density of magnons (small chemical potential $\mu$) the system
is mainly populated by particles with momenta around the two
dispersion minima at $\pm Q$, which at low energies can be interpreted
as two different bosonic flavors.
One can obtain the low-energy effective theory in the form of the
 Gross-Pitaevsky-type energy functional for two-component field (for the outline
of the derivation, see Appendix \ref{app:GP}):
\begin{eqnarray} 
\label{GP} 
\mathcal{H}_{\rm GP}&=&\int dx\,\Big\{ \sum_{a=1,2}\frac{|\nabla
\Phi^{a}|^{2}}{2m}
+\frac{1}{2}\Gamma_{11}^{\Lambda}(n_{1}^{2}+n_{2}^{2})\nonumber\\
&+&\Gamma_{12}^{\Lambda} n_{1}n_{2} -\mu(n_{1}+n_{2})\Big\}.
\end{eqnarray}
Here $\Phi^{1,2}$ are the   macroscopic 
bosonic fields that describe magnons with momenta $k$ lying within the
intervals $|k\pm Q|<\Lambda$ around the dispersion minima, $\Lambda$
is an infrared cutoff, and $n_{a}=|\Phi^{a}|^{2}$ are the
corresponding densities. 
Both bosonic species have the same effective mass
\begin{equation} 
\label{mass} 
m=\left(\left.\frac{\partial^{2}\varepsilon_{k}}{\partial k^{2}}\right|_{k=Q} \right)^{-1}=\frac{2}{SJ_2 (16-\beta^{2})}.
\end{equation} 
For the sake of clarity, the Planck constant
is set to unity. 

The macroscopic couplings $\Gamma_{ab}^{\Lambda}$ are given by
\begin{eqnarray} 
\label{Gamma-Lambda} 
\Gamma_{11}^{\Lambda}&=& \Gamma_{0}^{\Lambda}(Q,Q;0),\nonumber\\
\Gamma_{12}^{\Lambda}&=& \Gamma_{0}^{\Lambda}(-Q,Q;0)+\Gamma_{2Q}^{\Lambda}(-Q,Q;0)
\end{eqnarray}
where the vertex function $\Gamma_{q}^{\Lambda}(k,k';E)$ is the solution of the BS
equation (\ref{BS-gen}) with 
the  infrared cutoff $|p|>\Lambda$ introduced into the sum
over transferred momentum $p$.

As noticed first by Batyev and Braginskii,\cite{BatyevBraginskii84}
for a model with short-range interaction the solution of
(\ref{BS-gen}) 
can be expressed  in terms
 of a finite number of Fourier
harmonics in the transferred momentum. In our case, from the structure
of $V_{q}(k,k')$ it is easy to see
that $ \Gamma_{q}(k,k';E)$ generally contains harmonics
proportional to $1$, $\cos q$, $\sin q$, $\cos 2q$, and $\sin 2q$. The
integral equation thus is reduced to  a 
system of five linear equations that can be solved for general spin $S$,
thus no $1/S$ expansion is necessary.
For the purpose of finding only
the effective couplings $\Gamma_{11}$, $\Gamma_{12}$, the number of linear
equations in the general problem
outlined above can be reduced from five to three,
see Appendix \ref{app:BS-sym}.

The point $\Gamma_{11}=\Gamma_{12}$
corresponds to the enhanced $SU(2)$ symmetry at the level of the
effective low-energy theory.
For $\Gamma_{12}<\Gamma_{11}$,  the ground state of the gas contains an equal
density of the two particle species, and for  $\Gamma_{12}>\Gamma_{11}$
 just one
of the two species is present in the ground state. 
Recall that in our spin problem the total number of each
species is not separately fixed, in contrast to a typical setup for
atomic Bose mixtures. In a setup with fixed particle numbers the
ground state
at  $\Gamma_{12}<\Gamma_{11}$ is in the mixed phase, and
$\Gamma_{12}>\Gamma_{11}$ corresponds to phase separation. 
In the spin language, in two or three dimensions the state with only
one bosonic species 
corresponds to the helical magnetic order, while the state with an equal
density of the two species corresponds to the ``fan'' phase with coplanar
spin order.\cite{UedaTotsuka09} In one dimension, no long-range order
is possible, and the  state with only one species can be identified
with the vector chiral (VC) phase,\cite{KV05} while the two-species state corresponds
to the  two-component Tomonaga-Luttinger (TLL2)
liquid.\cite{FathLittlewood98,Okunishi+99}

The  approach that we have just described is well-known in the physics of magnetism in a
strong field
\cite{BatyevBraginskii84,Gluzman94,NikuniShiba95,JackeliZhitomirsky04,Okunishi98,Johnson86,UedaTotsuka09}
and is rather straightforward for three-dimensional systems where in
the limit $\Lambda\to0$ the couplings $\Gamma_{ab}^{\Lambda}$ are well
defined and acquire finite values.  In the case of low dimensions,
however, for $\Lambda\to 0$ the integral in Eq.~(\ref{BS-gen})
diverges at $p= 0$ for $(k,k')=(Q,Q)$, and for $(k,k')=(-Q,Q)$ there
are divergences both at $p=0$ and $p=2Q$.  Thus, the above scheme
needs a modification in the low-dimensional case.  The couplings
$\Gamma_{ab}^{\Lambda}$ can be interpreted as functions of the running
cutoff $\Lambda$ in the spirit of the renormalization group (RG)
approach.  The RG flow is then interrupted at a certain scale
$\Lambda=\Lambda_{*}$ that depends on the particle density. This
approach has proved to be quite successful for one-component Bose
gas.\cite{FisherHohenberg88,NelsonSeung89,KolomeiskyStraley92,Kolomeisky+00}
The focus of our study here is to extend this approach to the
two-component gas, which, as we have seen, is of special interest for
frustrated systems.  In this paper, we concentrate on the
one-dimensional case.

Solving the BS equation (\ref{BS-gen}), one 
can show that the expansion of
$\Gamma_{ab}^{\Lambda}$ as a series in $\Lambda$ has the following form:
\begin{eqnarray} 
\label{barG-series} 
\frac{1}{\Gamma_{11}^{\Lambda}}&=&\frac{m}{\pi\Lambda} +
\frac{1}{g_{11}}+O(\Lambda)+\ldots,\nonumber\\
\frac{1}{\Gamma_{12}^{\Lambda}}&=&\frac{m}{\pi\Lambda} +
\frac{1}{g_{12}}+O(\Lambda)+\ldots.
\end{eqnarray}
Note that the expansion starts with the term proportional to
$1/\Lambda$ that
turns out to be the same for both couplings.
Thus, for $\Lambda\to0$ both $\Gamma_{11}^{\Lambda}/\Lambda$ and $\Gamma_{12}^{\Lambda}/\Lambda$
flow to the same universal value, which reflects the tendency of the
RG flow to restore the $SU(2)$ symmetry for the two-component Bose
mixture.\cite{Kolezhuk10}

\begin{figure}[tb]
 \includegraphics[width=0.3\textwidth]{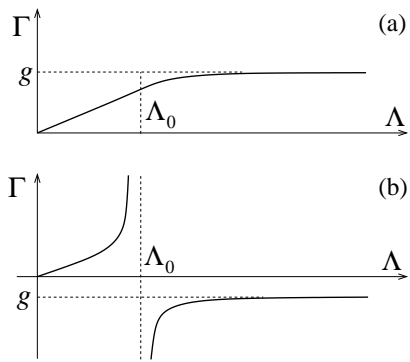}
 \caption{
\label{fig:Gamma-Lambda}
The typical behavior of the running coupling constant $\Gamma$ as a
function of the infrared cutoff $\Lambda$, in a \emph{continuum} model
of a Bose gas with a
bare contact  interaction $g$ (i.e., the two-body interaction
  potential is approximated by the delta-function $U(x)=g\delta(x)$): (a) for the case of
the repulsive 
coupling $g>0$, (b) for the case of attraction $g<0$.
}
\end{figure}

The parameters $g_{11}$, $g_{12}$, which 
determine the second term in the expansions
(\ref{barG-series}), have a special meaning: under certain conditions
they can be identified (see Appendices \ref{app:GP},\ref{app:1-comp}) with
the effective bare coupling constants of the \emph{continuum} two-component Bose gas with
a contact interaction $g_{11}\delta(x-x')$  between the
same species  (here $\delta(x)$ is the Dirac delta-function) and an
interaction of the form $g_{12}\delta(x-x')$ between different
species. Such a continuum model is essentially the 
two-component Lieb-Liniger model. 
The behavior of running couplings $\Gamma_{ab}^{\Lambda}$ in this problem is exactly given by the first two terms in
Eq.~(\ref{barG-series}).\cite{Kolezhuk10}
In the semiclassical limit $S\to\infty$, the parameters $g_{ab}$ are
given by simple expressions\cite{note-oldpaper}
\begin{eqnarray} 
\label{g-class} 
g_{11}&=&\beta^{2}/4+2(1+J_{2}^{z}+J_{1}^{z})=H_{s}/S,\nonumber\\
g_{12}&=&\frac{J_{2}^{z}\beta^{4}}{16}+ \beta^{2}(\frac{1}{2}+\frac{J_{1}^{z}}{4}-J_{2}^{z})+4(J_{2}^{z}+1)\,.
\end{eqnarray}

Typical behavior of $\Gamma_{ab}^{\Lambda}$ for the Lieb-Liniger model
is shown in
Fig.\ \ref{fig:Gamma-Lambda}. 
It is clear that $\Gamma_{ab}$ changes
its behavior on the characteristic scale $\Lambda_{0}\sim g_{ab}m$. For
the repulsive case ($g_{ab}>0$) this is simply the scale at which the
renormalized coupling $\Gamma_{ab}^{\Lambda}$ starts to deviate considerably
from its bare value, while for the case of attraction ($g_{ab}<0$) the
renormalized coupling has a pole at $\Lambda=\Lambda_{0}$ that 
technically limits
the RG flow and physically
indicates the existence of bound states with energy $E_{b}\sim \Lambda_{0}^{2}/2m$.

It is important to realize the following:
since we actually deal with a \emph{lattice} problem that has a
natural ultraviolet cutoff $\Lambda_{\rm UV}\sim 1$, the interpretation of
$g_{ab}$ as physical bare couplings makes sense only if
the  condition
\begin{equation} 
\label{appl-cont} 
|g_{11}|m \ll 1,\quad |g_{12}|m \ll 1
\end{equation}
is satisfied (see Appendix {\ref{app:1-comp}} for more details).  If the above condition of weak coupling is broken,
parameters $g_{ab}$ cannot be interpreted as physical bare
couplings, and only the fully renormalized interactions
\[
\Gamma_{ab}=\Gamma_{ab}^{\Lambda=\Lambda_{*}}
\]
retain their meaning as
effective low-energy coupling constants. For instance, if $g_{ab}<0$ but
$|g_{ab}|m \gtrsim 1$, there is no attraction in the lattice problem and
no bound states are present: directly looking at the behavior of
$\Gamma_{ab}^{\Lambda}$ for the lattice problem, one can convince oneself that there is no
pole in $\Lambda$.  In the strong coupling regime $|g_{ab}|m \gtrsim 1$
the sign of $g_{ab}$ merely determines the curvature of the
$\Gamma_{ab}^{\Lambda}$ dependence near $\Lambda=0$, but the
renormalized coupling itself remains positive.  The only physical meaning
of $g_{ab}$ in such a case is that they are connected to the
asymptotic phase shift of scattering states at  small transferred
momenta, see Sect.\ \ref{sec:2magnon} below.

The scale 
$\Lambda_{*}$, at which the RG flow has to be interrupted, can be easily fixed from the well-known exact result
for the single-component dilute Bose gas:\cite{Girardeau60} in the limit $\mu\to0$ the
particle density $n$ must behave as
\begin{equation} 
\label{1d-magcurve} 
n=(2m\mu/\pi^{2})^{1/2}\,.
\end{equation}
Indeed,  
in the phase with
only one bosonic species, which corresponds to
$\Gamma_{12}>\Gamma_{11}$, the density of particles obtained from the semiclassical description of Eq.\ (\ref{GP}) is
$n=\mu/\Gamma_{11}$, so in order to fullfil Eq.~(\ref{1d-magcurve}), one has to
assume that at $\mu\to 0$ the effective coupling 
behaves as $ \Gamma_{11} \to \pi\sqrt{\mu/(2m)}$ and  thus
\begin{equation} 
\label{Lambda*}
\Lambda_{*}=\sqrt{\mu m/2}.
\end{equation}
In the phase with equal densities of different species, which
corresponds to $\Gamma_{12}<\Gamma_{11}$, the total density at
the semiclassical level of Eq.\ (\ref{GP}) is
$n=2\mu/(\Gamma_{11}+\Gamma_{12})$. It follows from
Eq.\ (\ref{barG-series}) that for $\mu\to 0$ the behavior of the
density is described by the same expression (\ref{1d-magcurve}) as for
the one-component gas. Thus, at the transition from the vector chiral
phase to the two-component TL liquid, in the regime of a low magnon density there is no magnetization kink (cusp),
in agreement with numerical
results.\cite{OkunishiTonegawa03,Okunishi08,HM+09} The reason for that
is the effective fermionization of the magnon gas and formation of a
single Fermi sea at low densities, see the discussion below in Sect.\ \ref{sec:2magnon}.

From Eqs.\ (\ref{barG-series}) one can see that the transition point between
the TLL2 and VC phases, that is determined by the condition $\Gamma_{11}=\Gamma_{12}$,
can be detected by the 
\emph{crossing} of $g_{12}$ and
$g_{11}$ (i.e., a change of sign of $g_{12}-g_{11}$ that may happen if one of
the couplings goes through a pole does not correspond to a phase transition). We
will consider specific examples below in Sect.\ \ref{sec:chain-results}.

There is another way of calculating the renormalized interaction in
low dimensions,
which avoids introducing the infrared cutoff. In
Refs.\ \onlinecite{Lee+02,Morgan+02} it has been shown that a good
approximation to the
correct ``many-body'' expression for the effective four-point vertex
function (i.e., the two-particle scattering amplitude in the presence of a
finite particle density) at low energy  $\Gamma^{MB}_{q}(k,k';E\to0)$
can be obtained from the ``bare'' 
two-particle scattering amplitude  $\Gamma_{q}(k,k';E)$
 taken at a finite \emph{negative} energy $E=-E_{*}$ that is proportional
to the chemical potential $\mu$:
\begin{equation} 
\label{MB-2B} 
\Gamma^{MB}_{q}(k,k';E\to0) \approx \Gamma_{q}(k,k';-E_{*}),\quad
E_{*}=C\mu.
\end{equation}
Here the coefficient $C$ turns out to  depend on the dimension $d$ of the problem: for
$d=2$ one obtains $C=1$, and $C=\pi^{2}/8$ for $d=1$. 
The two-particle scattering amplitude  $\Gamma_{q}(k,k';E)$ is
precisely  the  solution of Eq.\ (\ref{BS-gen}) since $\mu=0$
means that no other particles are present except for the two that scatter off 
each other.

Following this scheme, we can define 
the energy-dependent coupling constants as
\begin{eqnarray} 
\label{Gamma-E} 
\Gamma_{11}(E)&=& \Gamma_{0}(Q,Q;E),\nonumber\\
\Gamma_{12}(E)&=& \Gamma_{0}(-Q,Q;E)+\Gamma_{2Q}(-Q,Q;E)
\end{eqnarray}
(note that the expressions on the rhs of Eq.\ (\ref{Gamma-E}) are now
calculated without any  infrared cutoff).
Solving the BS
equation  in the symmetrized form (\ref{sym-G11}),
(\ref{sym-G12}) for $\Gamma_{ab}(-E_{*})$, one 
can show that the expansion of
$\Gamma_{ab}^{-1}(-E_{*})$  in $E_{*}$ has the following structure:
\begin{eqnarray} 
\label{enG-series} 
\frac{1}{\Gamma_{11}(-E_{*})}&=&\left(\frac{m}{4E_{*}}\right)^{1/2} +
\frac{1}{g_{11}}+O(E_{*}^{1/2})+\ldots,\nonumber\\
\frac{1}{\Gamma_{12}(-E_{*})}&=&\left(\frac{m}{4E_{*}}\right)^{1/2} +
\frac{1}{g_{12}}+O(E_{*}^{1/2})+\ldots
\end{eqnarray}
The similarity with the expansion in the infrared cutoff, i.e., 
Eq.~(\ref{barG-series}), is obvious. 
Replacing  $E_{*}$ with its 1d
value $E_{*}=\pi^{2}\mu/8$,
one can readily convince oneself that  
\begin{equation} 
\label{en-bar-equiv} 
\Gamma_{ab}^{\Lambda_{*}} = \Gamma_{ab}(-E_{*}) +O(\mu^{3/2}),
\end{equation}
i.e., both regularization methods yield equivalent results in
the dilute limit $\mu\to0$. However, for practical purposes, the off-shell 
method is more convenient since it allows an easier calculation of all
integrals involved.


\section{Two-magnon problem in frustrated spin chains}
\label{sec:2magnon}

A general
two-magnon state with the total
quasi-momentum $K$ can be written in the form 
\[
 |\Psi_{K}\rangle =\sum_{n,r\ge 0} C_{r} e^{iK(n+\frac{r}{2})}
\frac{S^{-}_{n} S^{-}_{n+r}|F\rangle }{\langle F| S^{+}_{n+r} S^{+}_{n} \,\,   S^{-}_{n} S^{-}_{n+r}|F\rangle} ,
\]
where $|F\rangle$ is the fully polarized state.
The Schr\"odinger equation for the two-magnon problem leads to a
coupled system of equations for the amplitudes $C_{r}$, presented in Appendix
\ref{app:twomagnon}, that is a generalization of the one for the case
of an $SU(2)$ symmetric exchange.\cite{Arlego+11} Both scattering
states and bound states of two magnons can be obtained from the
solution of this system.  The bound states can be obtained along the
lines derived for an $SU(2)$ symmetric exchange.\cite{Arlego+11} Here
we discuss the solutions for the case of continuum states that
describe scattering of two magnons with momenta $k_{1}$, $k_{2}$.

Solving this scattering problem, we have to distinguish two cases. In the
first case
the problem is non-degenerate, meaning that for a given
total momentum $K=k_{1}+k_{2}$ and energy $E$ there is only one scattering state. 
This will be 
realized  when the momenta of the two magnons participating in
the scattering are in the vicinity of the same dispersion minimum, e.g.,
$k_1=Q+k$, and $k_2= Q-k$. In this non-degenerate case
the system (\ref{Sch}) can be solved by using
the following ansatz:
\begin{equation}
\label{SState}
 C_r=\cos \big(kr+\delta_{11}\big)+v e^{-{\kappa_0} r}, \,\, r \ge 1.
\end{equation}
This ansatz satisfies the recurrence relation in
Eqs.\ (\ref{Sch}), for $(r\geq3)$ if ${\kappa_0 }$ is a root of the following
equation: 
\begin{equation}
J_1\cos{\frac{K}{2}}( \cos{k} - \cosh{\kappa_0 } ) 
+ J_2\cos{K}(\cos{2k} - \cosh{2\kappa_0 })=0.
\end{equation}
Furthermore, to be a physically acceptable solution, the correct root should satisfy $\mbox{Re} [ \kappa_0 ]>0$.
The three unknown constants $C_0$, $v$, and $\delta_{11}$ are determined from
the first three equations of (\ref{Sch}).

To extract the scattering length $a_{11}$ from the phase shift
$\delta_{11}$, one can use the same relation as found in the
one-dimensional continuum model of a one-component Bose
gas interacting via a  short-range potential,
namely  $\delta_{11}(k)\to -\frac{\pi}{2}-    a_{11} k$ at $k\to 0$,
or, in a different form,
\begin{equation}
\label{scatlength11}
 a_{11}=\lim_{k\to 0} \big[\cot(\delta_{11})/k\big].
\end{equation}
If the short-range potential between two particles of mass $m$ with coordinates
$x$, $x'$ is approximated by the
contact potential of the form $U(x-x')=g_{11}\delta(x-x')$,  then the
scattering length $a_{11}$ is connected to the bare coupling $g_{11}$
as follows:\cite{Olshanii98}
\begin{equation}
\label{g11-a11} 
g_{11}=-2/(a_{11}m). 
\end{equation}

\begin{figure}[tb]
\includegraphics[width=7cm]{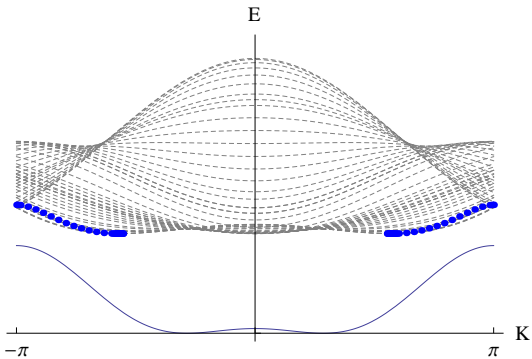}
\caption{
\label{fig:bound} (Color online).
Single magnon and two-magnon spectra for  the isotropic spin-1 chain with
$\beta=-2.5$ at the field $H>H_{s}$, as a function of the total
momentum $K$. The solid line shows the single-magnon dispersion with
minima at $\pm Q$.  One can clearly see degeneracies of the
scattering states in the vicinity of the lower bound of the
two-particle continuum for $K=0$ as well as in some high-energy regions.
A peculiar feature of the spectrum is the existence of a stable
two-magnon bound state inside the continuum of  scattering states
(this happens for any spin $S>\frac{1}{2}$ and $\beta<0$). The
procedure of finding two-magnon
bound states in isotropic ferromagnetic $S>\frac{1}{2}$ frustrated
chains  was discussed in detail in Ref.\ \onlinecite{Arlego+11}, and
here we include them for the sake of completeness. 
 }
\end{figure}

In the second
case, there is more than one scattering state for a given
energy and total momentum. This degeneracy stems from the
parity symmetry and the double-minimum structure of the single-magnon dispersion.
This case will be realized if the momenta of two scattering magnons are in the
vicinity of two different dispersion minima. For instance, the  state with the total momentum $K=0$
that corresponds to $k_1= Q+p$, $k_2= -Q-p$ (half relative momentum $k=Q+p$)
is degenerate with another scattering state with $k_{1}=Q-\tilde{p}$,
$k_{2}=-Q+\tilde{p}$ (half relative momentum $\tilde{k}=Q-\tilde{p}$).
This degeneracy is visible at the lower bound of the 
two-magnon scattering continuum around $K\simeq 0$, depicted in Fig.~\ref{fig:bound}.
Generally, $\tilde{p}\not= p$ because the shape
of the one-magnon dispersion curve is not
exactly symmetric  around the minima at $\pm Q$, but $\tilde{p}\to p$ at $p\to 0$. The  relative
momenta $k$, $\tilde{k}$ of two degenerate states with the same
total momentum $K$ are connected by the equation
\begin{equation}
 J_1\cos{\frac{K}{2}}( \cos{k} - \cos{\tilde k } ) 
+ J_2\cos{K}(\cos{2k} - \cos{2\tilde  k })=0.
\end{equation}

For $K=0$ the solution of the above equation takes the form
\begin{equation}
\tilde  k= \arccos(2\cos Q-\cos k),
\end{equation}
from which, in particular, it follows that for $p\equiv(k-Q)\to 0$ one
has $\tilde{p}\equiv (Q-\tilde{k})=p+O(p^{2})$.

The  Schr\"odinger equation for the scattering states  Eq.~(\ref{Sch}) 
is in this case  solved by the following ansatz:
\begin{equation}
\label{SStateDegenerate1}
 C_r=\cos \big(kr+\alpha(k) \big)+v(k) \sin(\tilde kr), \,\, r \ge 1.
\end{equation}
From the form of the scattering state Eq.~(\ref{SStateDegenerate1}) it
follows that if one prepares the incoming state of two magnons at $t\to
-\infty$ with relative momentum $k$, the outgoing state at $t\to
+\infty$ will be a superposition of states with relative momenta $k$
and $\tilde k$.

To determine the three  unknown quantities $C_0, \alpha$, and $v$ we again use the first three
equations of Eq.~(\ref{Sch}).  In Appendix \ref{app:interscattering} we
show how to extract 
 the interspecies scattering phase shift
\begin{equation}
\label{correctinterphase}
\delta_{12}= -\arccos{\frac{\cos{ \alpha }}{  \sqrt{\cos^2{\alpha}+(v+\sin{\alpha})^2}  }}
\end{equation}
from the scattering state
Eq.~(\ref{SStateDegenerate1}).
The scattering length is then obtained in analogy to the case of the intraspecies scattering,
\begin{equation}
a_{12}=\lim_{p\to 0} \big[ \cot(\delta_{12})/p\big]
\end{equation}
and the interspecies coupling is given by a formula completely
similar to Eq.~(\ref{g11-a11}):
\begin{equation}
\label{g12-a12}
 g_{12}=-2/\big(m a_{12}\big).
\end{equation}
It is worth noting that scattering states vanish when $p\to0$, i.e.,
when the magnons' momenta tend to the dispersion minima $k_{1}\to Q$,
$k_{2}\to -Q$. This is a consequence of fermionization at low
densities and indicates that different ``species'' in fact form a
\emph{single Fermi sea} at low energies.
 The existence of a single
Fermi sea is the reason for the absence of a magnetization cusp at the
transition between the vector chiral and TLL2 phases.\cite{note-1sea} 
It is worth noting that a similar phenomenon
is known in the one-dimensional Hubbard model: in the
limit of low electron density the electrons with different spin form
a common Fermi sea.\cite{1D-Hubbard-book}

\begin{figure}[tb]
\includegraphics[width=7.0cm]{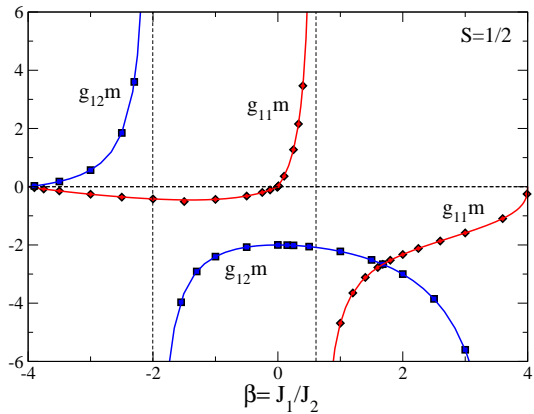}
\caption{
\label{fig:gm-s12}  (Color online).
Bare coupling constants for $S=1/2$: solid lines correspond to the
calculation based on the regularized Bethe-Salpeter equation, while symbols show
the results extracted from the solution of two-magnon scattering problem. In the region $\beta>0.614$, the
intraspecies coupling $g_{11}$ enters the super-Tonks regime.  The interspecies coupling
$g_{12}$ enters the super-Tonks regime in the region $\beta>-2$.
}
\end{figure}

The resulting coupling constants for $S=\frac{1}{2}$ chain are shown in
Fig.\ \ref{fig:gm-s12}. One can see that on the antiferromagnetic side
($J_1>0$) the scattering length $a_{11}$ is typically small (of the order of the
lattice constant) and goes through zero (i.e., $g_{11}$ has a simple
pole). Similarly, on the ferromagnetic side there is a pole in $g_{12}$.
In the previous section, we have seen that if the condition
Eq.~(\ref{appl-cont}) is not satisfied, a mapping to the effective
continuum model of the Lieb-Liniger type (with contact interactions) is problematic.
In relation to that,  we would like to discuss  the scattering length for an
unfrustrated ($J_{2}=J_{2}^{z}=0$) spin-$\frac{1}{2}$ $XXZ$ chain, which shows a similar behavior.
Setting $J_{1}=1$, one can write the analytical expression for this scattering length as
\begin{equation}
\label{a-XXZ}
a_{XXZ}=J_{1}^{z}/(1+J_{1}^{z}) \,.
\end{equation}
Close to the ferromagnetic point $J_{1}^z=-1$, it shows resonant behavior and this
allows one to develop the effective Lieb-Liniger model as the continuum
limit in this region.\cite{Seel+07} 
At $J_{1}^z=0$, however, $a_{XXZ}$ goes through zero and this
can be interpreted as follows: the continuum limit of the $XXZ$
spin-$\frac{1}{2}$ chain for $J_{1}^{z}>0$ does not fit into the description in terms of the repulsive
Lieb-Liniger model, rather, it is a super Tonks gas
(sTG)\cite{Astrakharchik-superTonks} with the Luttinger parameter $K<1$.
Including now the second neighbor interaction $J_{2}^{z}$ (but not the
hopping $J_{2}$), one obtains that at $J_{1}^{z}\to \infty$ and $J_{2}^{z}\to\infty$, the scattering length goes to $2$.
Thus, it  seems that in the regime of strong coupling $|g|m\gg1$ 
the scattering length has a different physical meaning, namely it
can be interpreted as the excluded volume of a particle. 
What is the correct continuum limit in this case,
however, remains a complicated issue.

\section{Analytical predictions for frustrated spin-$S$ chains}
\label{sec:chain-results}

\begin{figure}[tb]
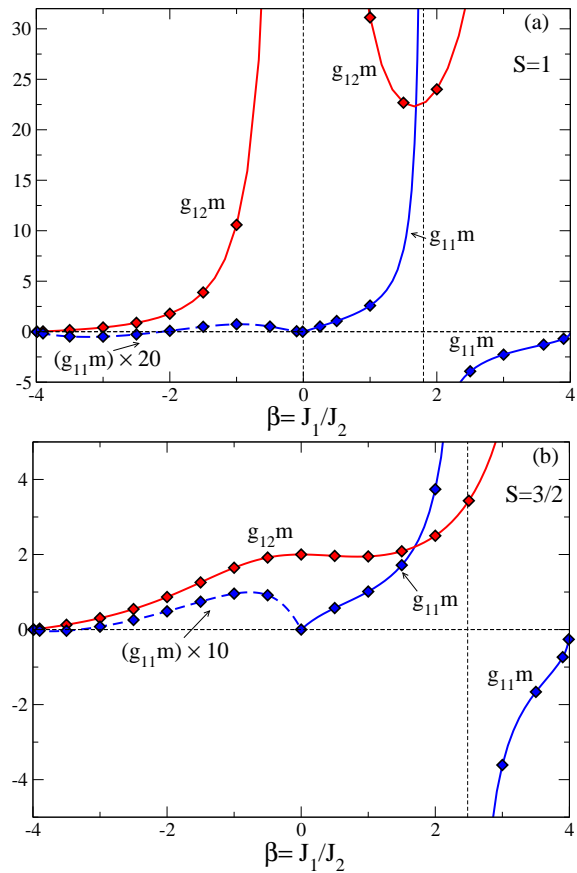

\includegraphics[width=0.42\textwidth]{gm-s1.eps}

\includegraphics[width=0.42\textwidth]{gm-s32.eps}
\caption{
\label{fig:gm-s1-s32}
(Color online). Bare coupling constants for isotropic frustrated spin
chains with (a) $S=1$ and (b) $S=\frac{3}{2}$: solid lines correspond to the
calculation based on the regularized Bethe-Salpeter equation, while symbols show
the results extracted from the solution of two-magnon scattering
problem. 
}
\end{figure}

The formalism developed in Sect.\ \ref{sec:efftheory} is rather
general and applies to any lattice Bose gas with degenerate
inequivalent dispersion minima. Let us look in more detail at the
predictions of our theory for the frustrated spin chain model
defined by the Hamiltonian (\ref{hamS}).

Figures \ref{fig:gm-s12}-\ref{fig:gm-s1-s32} show the values of the ``bare
couplings'' $g_{11}$ and $g_{12}$ for
 frustrated spin-$S$ chains with isotropic interactions
(i.e., $J_{1}^{z}=J_{1}$, $J_{2}^{z}=J_{2}$)
with $S=\frac{1}{2}$, $1$, and $\frac{3}{2}$ as functions of
the frustration parameter $\beta=J_{1}/J_{2}$. One can see that even for the lowest spin values
$S=\frac{1}{2}$ and $1$, there is an excellent agreement
between the approach based on the regularized Bethe-Salpeter equation
and the  results extracted from the direct solution of the two-magnon scattering problem.

\subsection{Ferromagnetic frustrated isotropic chains ($J_{1}<0$, $\Delta=1$)}

On the ferromagnetic side ($\beta<0$) the intraspecies coupling $g_{11}$
remains small enough for its interpretation as the bare coupling
of the effective continuum Lieb-Liniger model to be justified
(strictly speaking, this
is true for $S\geq 1$ where $|g_{11}|m\ll 1$, while for
$S=\frac{1}{2}$ the coupling values are higher, $|g_{11}|m\lesssim 0.5$).
For $S=\frac{1}{2}$,
this coupling remains negative in the entire range of $-4<\beta<0$,
which indicates the presence of bound states and is consistent
with 
known numerical results.\cite{Hikihara+08,Sudan+09,note-s12fm}

\begin{figure}[tb]
\includegraphics[width=7.0cm]{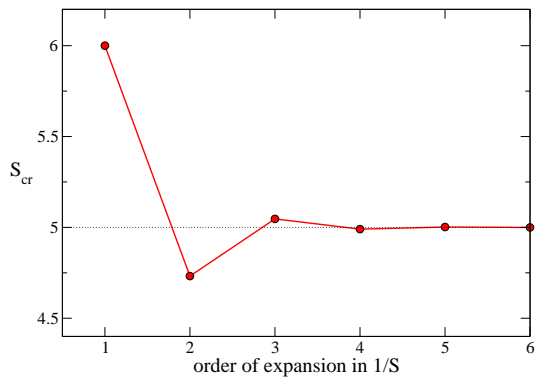}
\caption{
\label{fig:Scr} (Color online).
The critical value  $S_{\rm cr}$ of the spin that marks the disappearance of
the metamagnetic phase, as a function of the order $n$ of expansion in $1/S$.
The exact value is $S_{\rm cr}=5$, while the leading-term calculation
presented in Ref.\ \protect\onlinecite{Arlego+11} gives $S_{\rm cr}=6$.
}
\end{figure}

For $S=1$, the
intraspecies coupling $g_{11}$ becomes negative in the region $-4<\beta<\beta_{f}\simeq
-2.11$ (see
Fig.\ \ref{fig:gm-s1-s32}a), which signals the appearance of bound
magnon states. As shown previously in Ref.\ \onlinecite{Arlego+11} by means of
an $1/S$ expansion and numerical analysis, 
this also leads to metamagnetism (a finite  jump in the magnetization
at the saturation field).
The finite value of this jump can be understood in our approach by noticing the fact that
the full renormalized effective coupling $\Gamma_{11}^{\Lambda_{*}}$
depends on the particle density $n=1-M$, where $M$ is the
magnetization in units of saturation, and for $\beta<\beta_{f}$ the
coupling $\Gamma_{11}^{\Lambda_{*}}$ becomes positive at $M$ below some
critical value which determines the magnitude of the jump in $M$ that
is necessary to make the system stable again. 
The values of $M$ that are spanned by the jump correspond to states
with magnons collapsed into a single ``drop''. Such states have
negative magnetic susceptibility and are avoided in the ``grand
canonical'' setup when one fixes the external magnetic field $H$.

The numerical results of Ref.~\onlinecite{Arlego+11} also suggest that
the collapsed state has a finite vector chirality.  This is easy to understand by
observing that the interspecies interaction remains repulsive in the
entire ferromagnetic region $-4<\beta<0$.  Thus, it is energetically
favorable to form a ``drop'' involving only one magnon
species. As a side note, in this case $g_{11}$ can be directly
interpreted as the bare coupling of the effective continuum theory,
even though the interspecies coupling $g_{12}$ remains large
$|g_{12}|m\gtrsim 1$. This can be done because the system is
effectively in a one-component state.

\begin{figure}[tb]
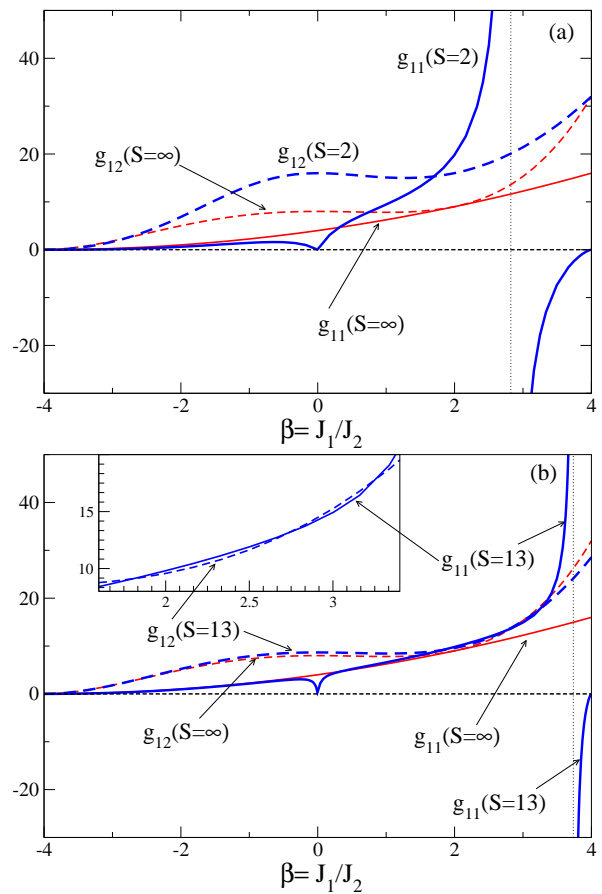

\includegraphics[width=78mm]{g12-g11-a.eps}

\includegraphics[width=78mm]{g12-g11-b.eps}

\caption{
\label{fig:g-class} (Color online).
Typical behavior of the ``bare couplings'' $g_{11}$ and $g_{12}$ for
isotropic spin chains with small and large spin $S$ as functions of the
frustration parameter $\beta$, compared to the semiclassical, $S\to 
\infty$ result (shown with thin solid line for $g_{11}$ and thin
dashed line for $g_{12}$): (a) for small $S$, there
is a single crossing of $g_{11}$ and $g_{12}$ curves, while (b) for
large $S\geq 12$, those curves cross three times (see the inset).
}
\end{figure}

Frustrated ferromagnetic chains with spins up to $S=\frac{9}{2}$
behave qualitatively in the same way as described above for
$S=1$. The interspecies ``bare coupling'' $g_{12}$ goes through a
simple pole
only for $S=\frac{1}{2}$ (for $S=1$ it goes through the pole of order
2 at $\beta=0$), so for $S\geq 1$
it becomes positive everywhere (see Figs.\ \ref{fig:gm-s12}-\ref{fig:gm-s1-s32},\ref{fig:g-class}).
The intraspecies coupling $g_{11}$ remains small and vanishes at $\beta=\beta_{f}$.
Upon increasing $S$, the
transition point $\beta_{f}$ quickly moves very close to $\beta=-4$
and  disappears for $S\geq 5$. This is clear from the behavior of
$g_{11}$ at $\beta\to -4$:
\begin{equation} 
\label{g11-at-4} 
g_{11}=\frac{S-5}{4(S+1)}(\beta+4)^{2}+O\left( (\beta+4)^{5/2}\right).
\end{equation}
Thus, $S_{\rm cr}=5$ is the critical value of spin where the
metamagnetic behavior vanishes in isotropic chains (for $S>5$, it exists
only in the presence of an easy-axis anisotropy).
We would like to note that Ref.\ \onlinecite{Arlego+11} reported a
slightly different value of $S_{\rm cr}=6$; this discrepancy is due to
the fact that Ref.\ \onlinecite{Arlego+11} used just the leading term
in the large-$S$ expansion, while our present approach is exact to all orders in
$1/S$. Figure \ref{fig:Scr} illustrates the behavior of $S_{\rm cr}$
as a function of the order of the $1/S$ expansion.

\begin{figure}[tb]
\includegraphics[width=70mm]{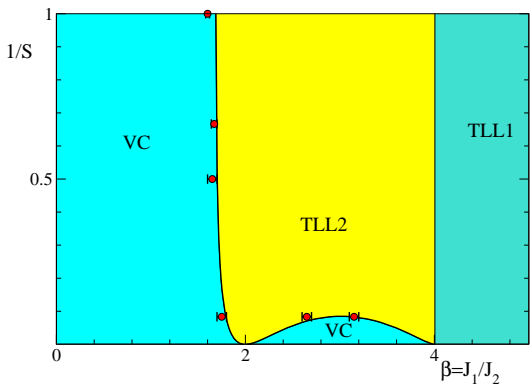}
\caption{
\label{fig:diag-S-beta} (Color online).
Phases of isotropic  frustrated antiferromagnetic
($J_{1}>0$) spin-$S$ chains  just below the  saturation field, as
predicted by our effective two-component Bose gas theory. The region
$\beta>4$ corresponds to a one-component TL
liquid phase (TLL1). 
If one varies $\beta$ at fixed $S$, there is a 
 single transition between the two-component
TL liquid (TLL2) and the vector chiral (VC) phase for $S<12$, 
 while 
for $S\geq 12$ one encounters three consecutive TLL2-VC transitions.
Symbols correspond to the transition points extracted from our
numerical analysis of chirality correlators, see
Section \ref{sec:numerics} for details.
}
\end{figure}

\subsection{Antiferromagnetic frustrated isotropic chains ($J_{1}>0$, $\Delta=1$)}

On the antiferromagnetic side $0<\beta<4$, the prominent feature of isotropic
chains is that for realistic (low) values of the spin $S$ the condition Eq.~(\ref{appl-cont}) is not
satisfied, as can be seen from
Figs.\ \ref{fig:gm-s12}-\ref{fig:gm-s1-s32}. 
Moreover,
$g_{11}$ changes its sign through a pole in this interval of $\beta$, for any value of $S$. As
discussed above, in this regime the system cannot be mapped to an
effective continuum Lieb-Liniger-type  model; however, a mapping to
the effective classical Gross-Pitaevskii model (\ref{GP}) is still possible, and
the corresponding effective couplings $\Gamma_{11}$ and $\Gamma_{12}$
remain positive in the entire antiferromagnetic region.
For large
$S$, the mass $m\propto S^{-1}$ becomes small (except near the two
Lifshitz points $\beta=\pm 4$) which ensures that in the limit
$S\to\infty$ the condition
Eq.~(\ref{appl-cont}) is again satisfied everywhere except for small regions around
the pole and the antiferromagnetic Lifshitz point.

Figure \ref{fig:g-class} shows the
typical behavior of the ``bare couplings'' $g_{11}$ and $g_{12}$ for
isotropic spin chains with small and large spin $S$, compared to the semiclassical $S\to 
\infty$ result. In the $S\to\infty$ limit, $g_{12}\geq g_{11}$
everywhere, and $g_{12}=g_{11}$ only at two points $\beta=-4$ (the
ferromagnetic Lifshitz point) and $\beta=2$ (the Majumdar-Ghosh point).\cite{note-oldpaper}
For small $S$, there
is only one crossing of $g_{11}$ and $g_{12}$  at some
$\beta=\beta_{1}<2$, indicating
a transition between the VC phase that exists in
the interval $0<\beta<\beta_{1}$ and the 
TLL2 phase that occupies the region $\beta_{1}<\beta<4$.
With increasing $S$, the transition point $\beta_{1}$ moves toward the
Majumdar-Ghosh point $\beta=2$, see Fig.\ \ref{fig:diag-S-beta} and
Table \ref{tab:Jcr}.
For very large $S\geq 12$, two more crossings 
of the $g_{11}$ and $g_{12}$ curves appear (see the inset in Fig.\ \ref{fig:g-class}b), and another piece of the
VC phase emerges in the region $2<\beta<4$, filling it completely as
$S$ goes to infinity
(see Fig.\ \ref{fig:diag-S-beta}).

\begin{table}
\begin{tabular}{|c|c|c|c|c|c|}
\hline
  $S$ & $1/2$  & 1  & $3/2$  & 2  &12 \\\hline
 $  \beta_{\rm TLL2-VC} $ & 1.682  & 1.688   &1.694  &1.702 & 1.80,
2.89, 3.13 \\\hline
\end{tabular}
\caption{\label{tab:Jcr} Critical values of the exchange coupling $\beta=J_{1}/J_{2}$
  for the transition from
  the VC to the TLL2 phase in isotropic frustrated antiferromagnetic  chains of spin $S$ at the
  saturation field,
 from the dilute Bose gas theory of the present work (cf. Figure
 \ref{fig:diag-S-beta}).
}
\end{table}

\begin{figure}[tb]
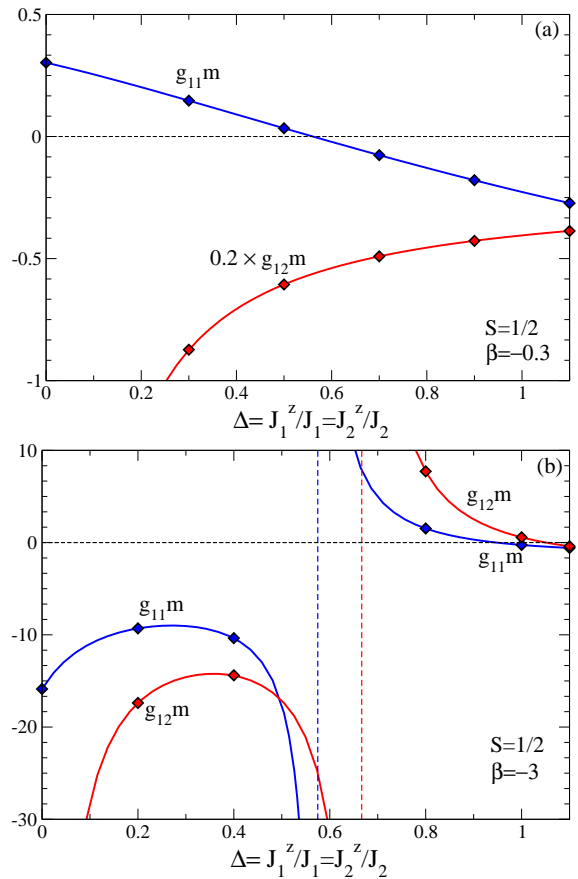

\includegraphics[width=0.42\textwidth]{gDelta-s12-b0.3.eps}

\includegraphics[width=0.42\textwidth]{gDelta-s12-b3.0.eps}
\caption{
\label{fig:gDelta} (Color online).
Bare coupling constants for $S=\frac{1}{2}$ anisotropic chain with
$J_{1}^{z}/J_{1}=J_{2}^{z}/J_{2}=\Delta$: (a) $J_{1}/J_{2}=-0.3$ and (b) $J_{1}/J_{2}=-3.0$.
}
\end{figure}

\subsection{Anisotropic chains ($\Delta\not=1$)}

Motivated by the availability of numerical results\cite{HM+09} for anisotropic
frustrated ferromagnetic chains with $S=1/2$, we would like to check
our theoretical predictions for anisotropic systems as well. Figure
\ref{fig:gDelta} shows the behavior of the ``bare couplings'' for 
$S=\frac{1}{2}$ anisotropic chains with two different  values of $\beta<0$, as  functions
of the anisotropy
$J_{1}^{z}/J_{1}=J_{2}^{z}/J_{2}=\Delta$. For $\beta=-0.3$, the
intraspecies coupling $g_{11}$ remains small and vanishes at
$\Delta\simeq 0.56$, while $g_{12}$ is negative and large
($|g_{12}|m\gtrsim 1$). 
This indicates a transition from the vector chiral phase at
$\Delta<0.56$ to a phase with bound states at $\Delta>0.56$. This agrees
favorably with the numerical results of Ref.~\onlinecite{HM+09} that reports a
transition between the chiral and nematic phases at $\Delta\simeq
0.6$.

\begin{figure}[tb]
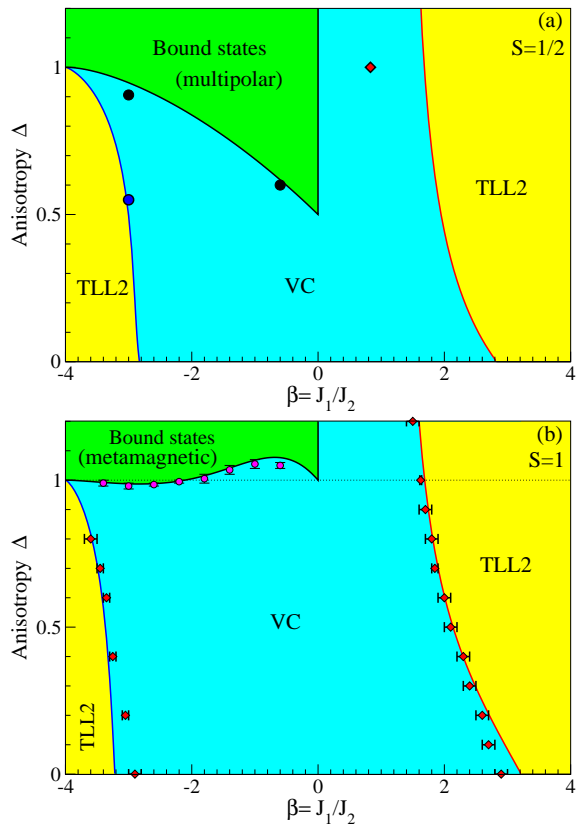

\includegraphics[width=0.42\textwidth]{s12-anis.eps}

\includegraphics[width=0.42\textwidth]{s1-anis-mod.eps}
\caption{
\label{fig:s-anis} (Color online).
Phase diagrams of  anisotropic
($J_{1}^{z}/J_{1}=J_{2}^{z}/J_{2}=\Delta$) frustrated chains with
(a) $S=\frac{1}{2}$  and (b) $S=1$ right
below the saturation field, derived from our dilute Bose gas
approach. The boundary between the two-component Tomonaga-Luttinger
liquid (TLL2) and the vector chiral (VC) phases has been extracted
from the condition $g_{11}=g_{12}$. At the boundary of the region labeled
``Bound states'' the intraspecies coupling $g_{11}$ changes its sign from positive
to negative (it turns out that $g_{11}$ always becomes negative before
$g_{12}$ does). The ``bound states''  region for $S=\frac{1}{2}$ in fact contains multiple phases
corresponding to bound states of a different number of
magnons,\protect\cite{Hikihara+08,Sudan+09} which is beyond the scope of the
present theory that can only detect the presence of two-magnon bound
states.  For $S=1$, the ``bound states'' region is
metamagnetic,\cite{Arlego+11} i.e., it exhibits a finite magnetization
jump immediately below the saturation field.
For $S=\frac{1}{2}$, circles correspond to the numerical data of
Ref.\ \onlinecite{HM+09}, and the diamond is the  location of the VC-TLL2
transition at $\Delta=1$ as found numerically in
Refs.\ \protect\onlinecite{Okunishi08,Hikihara+10}.
For $S=1$, symbols denote our DMRG results (see Section
\ref{sec:numerics} for details): diamonds and circles denote the phase transition
points extracted from the analysis of the chirality correlators and
the magnetization curves, respectively. 
}
\end{figure}

For $\beta=-3$, there is a crossing of $g_{11}$ and $g_{12}$ at
$\Delta\simeq 0.5$ (see Fig.\ \ref{fig:gDelta}b) that indicates a transition between the TLL2 phase
at $\Delta<0.5$ and the VC phase at $\Delta>0.5$, nicely fitting to
 the numerical data of Ref.\ \onlinecite{HM+09}. Both
$g_{11}$ and $g_{12}$ go through poles and become positive
(which does not change the phase since in the course of this change one
always has $\Gamma_{12}>\Gamma_{11}$), then they
become small and positive around $\Delta=1$, and $g_{11}$ crosses zero
at $\Delta\simeq 0.947$, indicating a transition from the VC phase to
a phase with bound states ($g_{12}$ changes its sign later, at about
$\Delta=1.05$). Again, this is in  good agreement with the
numerical study Ref.~\onlinecite{HM+09} that observes a transition 
around $\Delta=0.95$.

Figure \ref{fig:s-anis} shows the predicted phase diagram of 
anisotropic $S=\frac12$ and $S=1$ chains immediately below the
saturation field in the $(\beta,\Delta)$ plane. One can see that with
increasing $S$ the TLL2 phase shrinks and the phase with bound states
is pushed toward higher $\Delta$ (for $S>5$ it lies entirely in the
easy-axis region $\Delta>1$). The vector chiral phase dominates the
phase diagram.  
One can conclude that the present theory reliably
detects the transition points in anisotropic ferromagnetic chains even
at such  low values of the spin as $S=\frac{1}{2}$. It is worth noting that the
transition point between the TLL2 and VC phases in the $\beta=-3$, $S=\frac{1}{2}$
chain comes out correctly despite the fact that both $|g_{12}|m$ and
$|g_{11}|m$ are large at the transition. 
On the antiferromagnetic
side, the only numerical data available in the literature is for the
isotropic $S=\frac{1}{2}$
chain.\cite{Okunishi08,Hikihara+10} In this case, for a reason that
is not yet clear, the agreement between the theory and numerics is
considerably worse.


\section{Numerical analysis}
\label{sec:numerics}

To verify our theoretical predictions, we have performed extensive
density matrix renormalization group (DMRG) simulations.
For
a detailed description of the DMRG technique,\cite{white92} we refer the reader to
the review Ref.\ \onlinecite{schollwoeck05}. We focus on two quantities: First, we calculate the
central charge from the scaling behavior of the von Neumann entropy,
which allows us to distinguish one- from two-component phases. Second,
we compute the vector chirality order parameter from the analysis of
chiral correlation functions, to identify those one-component phases that
have long-range vector chiral order. The goal is to provide a numerical check of 
the phase diagrams for isotropic and anisotropic frustrated chains, Figs.~\ref{fig:diag-S-beta} and \ref{fig:s-anis}, respectively.
In order to identify the region with bound states in anisotropic chains, we calculate 
magnetization curves.
We have studied systems of different sizes $L$ ranging from $L=50$ to $L=400$, both with periodic and
open boundary conditions, typically keeping $600$ to $1500$ states in most calculations.

\subsection{Central charge}

Our theory suggests that close to
the saturation field, in antiferromagnetic frustrated chains with
spin values of $1/2 < S < 12$, one encounters two phase
transitions  when increasing $\beta$ from zero: (i) from 
the VC phase into the TLL2 phase and (ii) from the TLL2 phase
into the TLL1 phase (compare Fig.~\ref{fig:diag-S-beta}).  The TLL2
phase has central charge $c=2$, while the VC and the TLL1 phases have
$c=1$, suggesting that a calculation of $c$ can be helpful in
determining the phase boundaries.

To accomplish that, we exploit that the von Neumann entropy $S_{vN}$
can be obtained as 
a by-product of any DMRG calculation. Furthermore,  from the finite-size
scaling of $S_{vN}$, one can extract the central charge.
The von-Neumann entropy is defined as
\begin{equation}
S_{vN,L}(l)=- \mbox{tr}(\rho_l \ln \rho_{l}) \, ,
\end{equation}
where $\rho_l$ is the reduced density matrix for a subsystem of length
$l$ embedded in a chain of a finite length $L$. 
In a gapless state that is conformally invariant, the $l$ and $L$
dependence of the von Neumann entropy
is given by \cite{vidal03,calabrese}
\begin{equation}
S_{vN,L}(l) =\frac{c}{3} \ln\left\lbrack \frac{L}{\pi} \sin\left (\frac{\pi}{L} l\right )\right\rbrack+g\,,
\label{eq:cch}
\end{equation}
which is valid for systems with periodic boundary conditions
(PBC). Here $g$ is a non-universal constant that depends on the
magnetization $M$. As
DMRG directly accesses the eigenvalues of these reduced density matrices, \cite{white92,schollwoeck05}
it is straightforward to measure $S_{vN,L/2}(l)$ with this method. In
the case of open boundary conditions (OBC), the von Neumann entropy is
given by one half of the formula  Eq.~(\ref{eq:cch}), where besides $g$, there are also 
  contribution that oscillates as a function of $l$.

\begin{figure}[tb]
\includegraphics[width=0.42\textwidth]{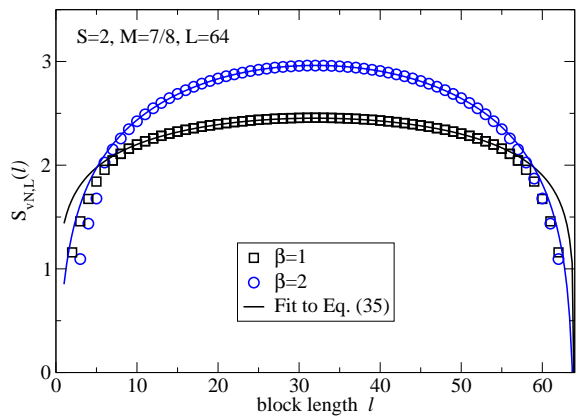}
\caption{
\label{fig:svn} (Color online).
The von Neumann entropy $S_{vN,L}(l)$ vs block length $l$ for $S=2$ at $M=7/8$ for
$\beta=2$ (circles) and $\beta=1$ (squares). The solid lines are fits to 
Eq.~\eqref{eq:cch}, resulting in $c\approx 1.01$ and $c=2.10$, respectively (the last 10
data points were excluded from the fits). Symbols are DMRG data for $m=1500$ states. 
}
\end{figure}

Typical DMRG results for the von Neumann entropy  of chains with $L=64$ spins and
PBC are presented in Fig.~\ref{fig:svn} for $S=2$, together with a fit
of Eq.~\eqref{eq:cch} to the numerical data (lines).  These fits
result in $c\approx 1$ for $\beta=1$ and $c\approx 2.1$ for $\beta=2$
at $M=7/8$.  
As it turns out, there
are strong finite-size effects in the estimation of $c$ from fits to
Eq.~\eqref{eq:cch},  in particular at high fields close to saturation
which are our primary interest. 
Nishimoto\cite{nishimoto11} has recently suggested
to calculate $c$ from the following formula:
\begin{equation}
c=\frac{3\lbrack S_{vN,L}(L/2-1)
  -S_{vN,L}(L/2)\rbrack}{\mbox{ln}\lbrack \cos(\pi/L)\rbrack}\, ,
\label{eq:nish}
\end{equation}
which turns out to converge much faster with the system size. However,
the numerical data needs to be sufficiently accurate for
Eq.~\eqref{eq:nish} to yield reliable results, and we found that one has to push the
number of states in a typical DMRG calculation to about
$m\sim 1500$ to achieve good accuracy. 
For instance, for the
parameter of Fig.~\ref{fig:svn}, Eq.~\eqref{eq:nish} results in
 $c=1.005$ for
$\beta=1$.  
We used Eq.~\eqref{eq:nish} to extract $c$ whenever the
DMRG calculation proved to be sufficiently accurate, and resorted to
fits to Eq.~\eqref{eq:cch} (or its OBC counterpart) otherwise.

\subsection{Chirality order parameter}

\begin{figure}[tb]
\includegraphics[width=0.42\textwidth]{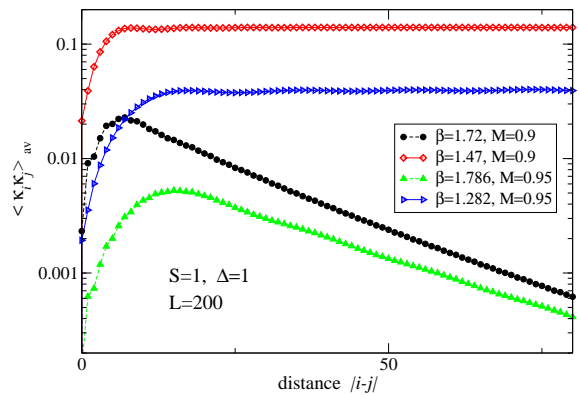}
\caption{
\label{fig:kappa-corr} (Color online).
Typical vector chirality correlation functions for an isotropic  $S=1$ 
chain of $L=200$ spins.
}
\end{figure}

To study the behavior of the vector chirality, we have calculated
correlators of the form
\begin{equation} 
\label{chir-corr}
 \langle\kappa_i\kappa_j \rangle, \quad \kappa_i= \left[\vec{S}_i \times \vec{S}_{i+1} \right]^z.
\end{equation}
 The chiral order parameter $\kappa$ has been extracted as 
\begin{equation}
\label{squarechirality}
\kappa^{2}= \lim_{|i-j|\gg 1} \langle\kappa_i\kappa_j \rangle_{\rm av},
\end{equation}
where the index ``av'' means that correlators were averaged over pairs of sites $(i,j)$ with the
same $|i-j|$, while care was taken to stay ``in the bulk'', i.e., $i$
and $j$ were kept sufficiently far away from the ends of the
chain. Figure \ref{fig:kappa-corr} shows typical vector chirality
correlation functions for a $S=1$ isotropic frustrated 
chain of  length $L=200$.

\subsection{Numerical results: Isotropic chains}

\begin{figure}[tb]
\includegraphics[width=0.42\textwidth]{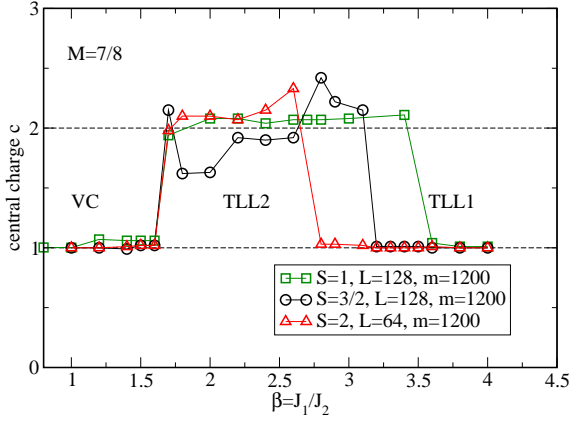}
\caption{
\label{fig:ccharge} (Color online).
Central charge at $M=7/8$ for $S=1$, $3/2$, and $2$, as extracted
from Eq.~\eqref{eq:nish} [The data points in the $c\sim 2$ region, i.e., $1.5<\beta<2.5$ were extracted from direct fits of
Eq.~\eqref{eq:cch} to the DMRG results for $S_{vN,L}(l)$].   }
\end{figure}

Our results for the central charge of isotropic frustrated chains are
shown in Fig.~\ref{fig:ccharge}. 
We have collected data 
at fixed magnetization $M$ for $S=1$, $\frac{3}{2}$, $2$. From the figure we clearly identify two phase transitions:
(i) at a critical, $S$-dependent $\beta \gtrsim 3$ from  TLL1 ($c=1$) phase into the TLL2 phase ($c=2$) and (ii)
at $\beta\gtrsim 1.6 $,  from the TLL2 phase into another $c=1$
phase. Based on our theory,
we expect this latter phase to have vector chiral order (which is
verified by calculating the chiral order parameter, see below).

The position of the phase transition from TLL2 into VC depends only
weakly on $S$, consistent with our theory (compare the values given in
Tab.~\ref{tab:Jcr}).  According to our theory, the phase transition
from TLL1 into TLL2 happens at approximately $\beta=4$, the Lifshitz
point, in the limit of $M\to 1$. Our numerical data for $M<1$,
however, shows that this transition sets in at a slightly smaller $
\beta \lesssim 3.5$ than what one would expect from
Fig.~\ref{fig:diag-S-beta}.  This suggests that the phase boundary
depends on $M$.  Indeed, for instance, in the case of $S=1$, this phase
boundary shifts to smaller values of $\beta$ as $M$ {\it decreases}
which can be deduced from the analysis of magnetization curves
presented in Ref.~\onlinecite{HM+07}, where this particular phase
transition shows up as a kink in $M(H)$. We have not systematically
studied the $M$ dependence as the DMRG data for $S_{vN,L}$ tends to
converge rather slowly at large $M$.   Our numerical results
for $c$ are in very good overall agreement with the theory.

\begin{figure}[tb]
\includegraphics[width=0.42\textwidth]{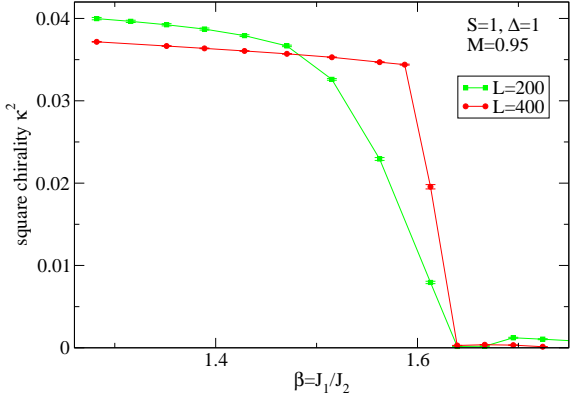}
\caption{
\label{fig:chiralitySpin1} (Color online).
Square of the vector chirality order parameter $\kappa^{2}$ for frustrated  isotropic
antiferromagnetic $S=1$ chain at
fixed value of the magnetization $M=95/100$, as
a function of the frustration parameter $\beta=J_{1}/J_{2}$.
}
\end{figure}

\begin{figure}[tb]
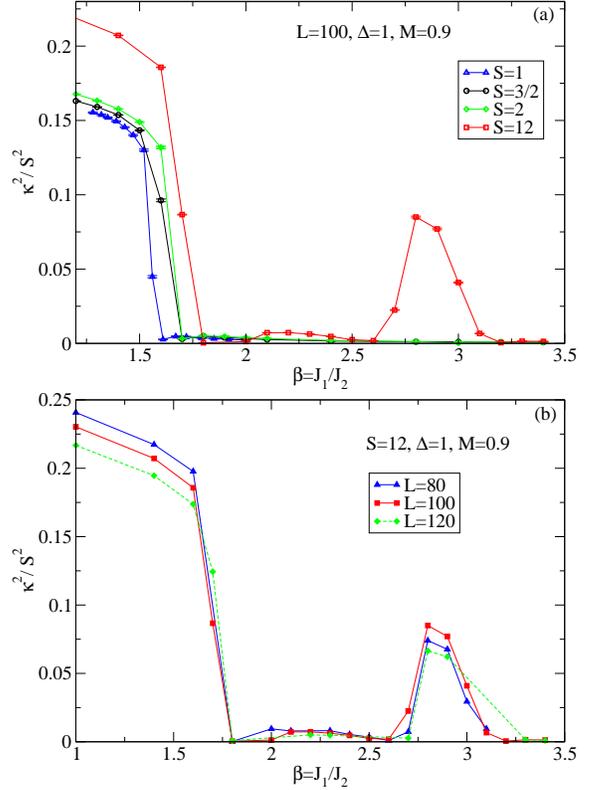

\includegraphics[width=0.42\textwidth]{sS_kappa_cut_vs_beta-mod.eps}

\includegraphics[width=0.42\textwidth]{s12_kappa_cut_vs_beta.eps}
\caption{
\label{fig:chiralitySpinS} (Color online).
Square of the vector chirality order parameter $\kappa^{2}$ for frustrated
isotropic antiferromagnetic spin-$S$ chains, at
fixed value of the magnetization $M=9/10$, as
a function of the frustration parameter $\beta=J_{1}/J_{2}$: (a) for
chains consisting of $L=100$ spins, at different values of $S$; (b)
for $S=12$ chains of different length $L$.
}
\end{figure}

We next verify that the $c=1$ phase at $\beta\lesssim 1.6$ is the VC phase,
by directly calculating the chiral order parameter as described
above; the results are shown in Figs.\ \ref{fig:chiralitySpin1} and
\ref{fig:chiralitySpinS}. One can clearly see from
Fig.\ \ref{fig:chiralitySpinS} that, in agreement with our analytical
predictions, the transition point $\beta$ from the vector chiral phase
to the two-component Tomonaga-Luttinger liquid increases slowly with
the increasing spin value $S$. Our numerical results for the
transition boundaries in isotropic frustrated chains are shown as
symbols in the phase diagram of Fig.\ \ref{fig:diag-S-beta}.

Figure \ref{fig:chiralitySpin1} illustrates that with the  increasing 
system size $L$, the VC-TLL2 transition becomes more steep. This
is consistent with the first-order character of this transition as
predicted by the theory (recall that the VC-TLL2 transition
corresponds to phase separation in the language of the effective
two-component Bose gas theory).

For high magnetization sectors it is possible to study large spins as
well and   results for $S=12$ are included in
Fig.\ \ref{fig:chiralitySpinS}. One should take those $S=12$ results with
some caution: our data for small systems
(consisting of 80, 100, and 120 sites) still show strong 
nontrivial size dependence (see Fig.\ \ref{fig:chiralitySpinS}b) that precludes infinite-size extrapolation,
 and we could not study larger systems because of the rapidly growing
dimension of the Hilbert space. 
Nevertheless, the resulting picture for $S=12$ qualitatively agrees with our analytical
prediction, namely, one can see the recurring TLL2-VC phase
transitions in accordance
with the phase diagram of Fig.\ \ref{fig:diag-S-beta}.

\subsection{Numerical results: Anisotropic $S=1$ chain}

\begin{figure}[tb]
\includegraphics[width=0.43\textwidth]{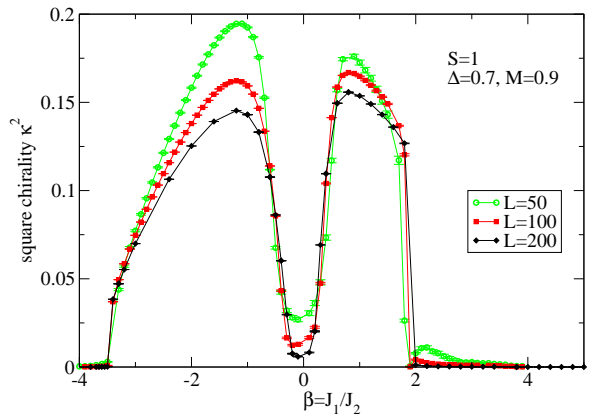}
\caption{
\label{fig:chiralityAnisotropic}
Square of the vector chirality order parameter $\kappa^{2}$ for frustrated 
antiferromagnetic $S=1$ chain for $\Delta=0.7$, at
fixed value of the magnetization $M=9/10$, as
a function of the frustration parameter $\beta=J_{1}/J_{2}$.
}
\end{figure}

We have studied numerically the phase diagram of the anisotropic
($J_{2}^{z}/J_{2}=J_{1}^{z}/J_{1}=\Delta$) frustrated spin-$1$  chain, to verify
the analytically predicted phase diagram of Fig.\ \ref{fig:s-anis}b.
Figure~\ref{fig:chiralityAnisotropic} shows the typical dependence of
the chiral order parameter $\kappa$ on the frustration parameter
$\beta$ for anisotropic   $S=1$
chain,  at fixed anisotropy $\Delta$ and magnetization $M$, for three different
system sizes. One can notice that finite-size effects are not always
monotonous, which complicates the extrapolation to infinite system size. 

\begin{figure}[tb]
\includegraphics[width=0.43\textwidth]{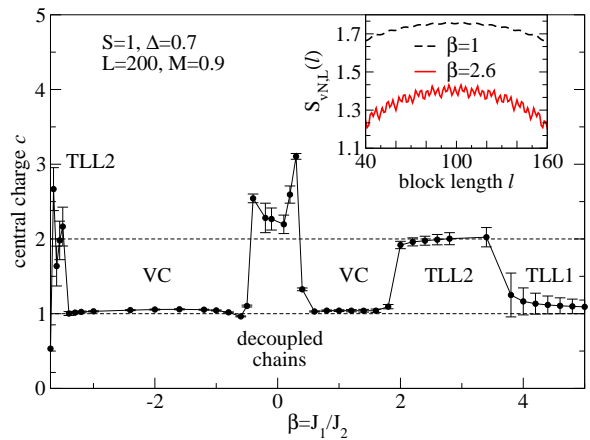}
\caption{
\label{fig:centralchargeAnisotropic}
Main panel: Central charge, obtained with open boundary conditions, for a frustrated
antiferromagnetic $S=1$  chain for $\Delta=0.7$, at
magnetization $M=9/10$, as
a function of the frustration parameter $\beta=J_{1}/J_{2}$.
Inset: von Neumann entropy $S_{vN,L}(l)$ versus $l$ at $\beta=1,2.6$.}
\end{figure}

In Fig.\ \ref{fig:centralchargeAnisotropic}, we present our results for
the dependence of the central charge on $\beta$, for the same
parameters as used for the chirality in Fig.\ \ref{fig:chiralityAnisotropic}. 
These data were obtained from systems with open boundary conditions 
for which we could obtain reliable data. The additional oscillatory 
contributions to $S_{vN}$ are evident from the inset of Fig.\ \ref{fig:chiralityAnisotropic}.
The behavior of the central charge is consistent with the results for the chirality:
Whenever $\kappa^2>0$, $c=1$. At both $\beta>2$ and $\beta\lesssim -3.5$, where 
$\kappa$ drops to zero, the central charge goes up to $c=2$, indicating phase transition from VC to TLL2 phase.

As for the regime of  almost decoupled chains $|\beta|\ll 1$, one should keep in mind that this is a
particularly difficult parameter region for  DMRG calculations, and
the  apparent deviation from the analytical predictions (according to
which $c=2$ is only realized for one point $\beta=0$) is an artefact of
the rapidly deteriorating accuracy at  $\beta\to 0$.

\begin{figure}[tb]
\includegraphics[width=0.43\textwidth]{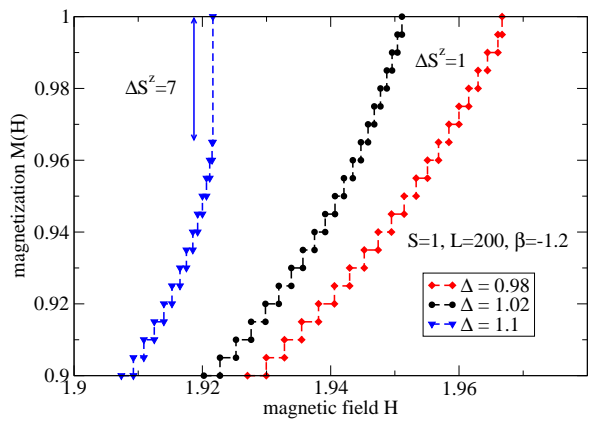}
\caption{
\label{fig:mh}
Typical magnetization curves for $S=1$ chain with 
$\beta=-1.2$, at different values of the anisotropy $\Delta$. For $\Delta=1.1$, the curve exhibits a metamagnetic jump
at the saturation field (similar jump for $\Delta=1$, for $-4<\beta<-2.11
$ has been already discussed in Ref.~\cite{Arlego+11}).
}
\end{figure}

In order to study the phase boundary between the VC and ``bound states'' (metamagnetic)
phase, we have also calculated magnetization curves, and looked
at the value $\Delta S^{z}$ of the steps by which the magnetization
changes. While inside the VC phase (where magnons interact
repulsively) the magnetization grows in  steps of $\Delta
S^{z}=1$ when increasing the magnetic field,  after crossing the metamagnetic
region boundary the height of steps in the magnetization curves
quickly changes to higher values $\Delta
S^{z}=2$, $3$, etc.
Typical results for $M(H)$ are shown in Fig.~\ref{fig:mh}.  
In this example ($S=1$, $\beta=-1.2$), we see metamagnetic behavior (a jump in $M(H)$) directly under saturation magnetization for $\Delta=1.1$, but not for $\Delta=1.02$ or $\Delta=0.98$.
The results for the onset of the metamagnetic behavior as a function of $\Delta$ and $\beta$
are included in Fig.~\ref{fig:s-anis}(b) (circles).

Our numerical results for the TLL2-VC phase boundary are shown by
symbols in Fig.\ \ref{fig:s-anis}b. The overall agreement
between the theory and numerics is very good.



\section{Summary}
\label{sec:summary}

We have developed the effective theory for low-density one-dimensional
lattice Bose gas with two degenerate dispersion minima. This study has
been motivated by the problem of the ground state phase transitions in
frustrated spin-$S$ chains subject to strong magnetic fields just
slightly lower than the saturation field, where such a system can be
viewed as a dilute lattice gas of magnons in the fully polarized
background. At low density of magnons they can be treated as bosonic
particles, and in a wide region of parameters the magnon dispersion
has two degenerate minima at inequivalent points $\pm Q$ in the
momentum space. 

The low-energy effective theory has the form of the Gross-Pitaevskii
functional for a two-component field. We have developed the scheme
that allows the calculation of renormalized effective interactions
$\Gamma_{11}$, $\Gamma_{12}$.  Applying this formalism to frustrated
spin-$S$ chains, we establish the magnetic phase diagram of isotropic
and anisotropic, ferro- and antiferromagnetic frustrated chains close
to saturation and study phase transitions between several nontrivial
states, including the two-component Tomonaga-Luttinger liquid (TLL2),
vector chiral (VC) phase, and phases with bound magnons. Our scheme
does not involve an $1/S$ expansion, and thus allows us to draw
quantitative predictions for arbitrary spin.

We complement our analytical predictions by a variety of numerical
results obtained by means of the density matrix renormalization group
technique.  Particularly, we compute the central charge as a function
of the frustration parameter close to saturation which allows us to
locate phase transitions between one- and two-component
states. Further, we directly compute chirality correlators in order to
identify those one-component states that have vector chiral order. In
addition, we study the magnetization curves to detect the presence of
magnon bound states.  These numerical results are in a good
quantitative agreement with our theoretical predictions for chains of
spin $S\geq 1$.  At the same time, in the case of frustrated antiferromagnetic
$S=\frac{1}{2}$ chains, the quantitative discrepancy between our
theoretical prediction and the numerical results by
Okunishi\cite{Okunishi08} and Hikihara \emph{et al.}\cite{Hikihara+10}
for the location of the TLL2/VC phase boundary is unexpectedly
large. Nevertheless, even for $S=\frac{1}{2}$ our theory still
correctly captures the topology of the phase diagram.

\begin{acknowledgments}

We thank H. Frahm for useful discussions.
A.K. gratefully acknowledges the hospitality of the
Institute for Theoretical Physics at the Leibniz University of
Hannover. This work has been supported by the cluster of excellence QUEST (Center for Quantum
Engineering and Space-Time Research). T.V. acknowledges SCOPES Grant IZ73Z0-128058.

\end{acknowledgments}


\appendix

\section{Low-energy effective theory and connection to the two-component
  Lieb-Liniger model}
\label{app:GP}

To obtain the low-energy effective theory for the Hamiltonian
(\ref{hamB}), 
we 
define the bosonic field $\psi=\sum_{k}b_{k}e^{ikx}$
with the Euclidean action given by 
$\mathcal{A}_{\psi}=\int d\tau\big\{ \int dx \psi^{*}\partial_{\tau}\psi
+\mathcal{H}[\psi]\big\}$, and divide the field $\psi$
into ``slow'' parts $\Phi^{1,2}$ and the ``fast'' part $\phi$:
\begin{eqnarray} 
\label{slowfast}
&& \psi=\sum_{\alpha=1,2}\Phi^{\alpha}e^{i\sigma_{\alpha}Qx}+\phi,\\
&& \Phi^{\alpha}=\sum_{|k|<\Lambda}
b_{\sigma_{\alpha}Q+k}e^{ikx},\quad  \sigma_{\alpha}=\begin{cases} 1,& \alpha=1 \\ -1, &
\alpha=2 \end{cases}
,\nonumber
\end{eqnarray}
where $\Lambda$ is the running cutoff.  Treating the ``slow'' part as
the background field, one can
integrate
over the fast modes 
$\int\mathcal{D}[\phi]e^{-\mathcal{A}_{\psi}}=e^{-\mathcal{A}_{\Phi}}$
and obtain the effective action  $\mathcal{A}_{\Phi}$ that is essentially
the generating functional of vertex functions\cite{Zinn-Justin-book} and can be
written as follows:
\begin{eqnarray} 
\label{Aeff} 
\mathcal{A}_{\Phi}&=&\Phi^{\alpha,*}_{k,\omega}
U_{k,\omega}^{(2),\alpha\beta}\Phi^{\beta}_{k,\omega} \\
&+&\frac{1}{2}\Phi_{k+q,\Omega+\omega}^{\alpha,*}\Phi^{\beta,*}_{k'-q,\Omega'-\omega}
U_{q,k,k',\Omega+\Omega'}^{(4),\alpha\beta}
\Phi^{\alpha}_{k,\Omega}\Phi^{\beta}_{ k', \Omega'}\nonumber\\
& +&(\text{higher-order terms in $\Phi$}).\nonumber
\end{eqnarray}
In the above equation, the summation over repeated indices is implied, and
the $\Lambda$-dependent coefficients $U^{(n)}$ can be expressed through the vertex
functions $\Gamma^{(n)}$
of the initial theory determined by the action $\mathcal{A}_{\psi}$:
\begin{eqnarray} 
\label{vertex}
 U^{(2),\alpha\beta}_{k,\omega}&=&\delta_{\alpha\beta}\Gamma^{(2)}(\sigma_{\alpha}
Q+k,\omega) ,\\
U^{(4),\alpha=\beta}_{q,k,k',\omega}&=&  \Gamma^{(4)}_{q}(\sigma_{\alpha}Q+k,\sigma_{\alpha}Q+k';\omega),\nonumber\\
U^{(4),\alpha\not=\beta}_{q,k,k',\omega}&=& \Gamma^{(4)}_{q}(\sigma_{\alpha}Q+k,\sigma_{\beta}Q+k';\omega)\nonumber\\
&+&\Gamma^{(4)}_{(\sigma_{\beta}-\sigma_{\alpha})Q+k'-k-q}(\sigma_{\alpha}Q+k,\sigma_{\beta}Q+k';\omega),
\nonumber
\end{eqnarray}
with the
restriction that sums in the internal lines of the
corresponding Feynman diagrams include only ``fast'' modes with momenta $p$
outside the regions $|p\pm Q|<\Lambda$. 

The two-point function $-\Gamma^{(2)}(k,\omega) $ is equal to the 
Green function $G(k,\omega)$. In the limit $\mu\to 0$, i.e., in the
absence of particles, the self-energy vanishes \cite{Uzunov81} and the
full propagator $G(k,\omega)$ is given by its bare value
$G^{(0)}(k,\omega)=(i\omega-\varepsilon_{k})^{-1}$. The four-point
vertex $\Gamma^{(4)}_{q}(k,k',\omega)\equiv
\Gamma_{q}^{\Lambda}(k,k';E=i\omega)$ 
 is determined by the Bethe-Salpeter equation, which in the limit
 $\mu\to0$ takes the simple form of Eq.\ (\ref{BS-gen}), with the
 infrared cutoff $\Lambda$ around the momenta $\pm Q$ introduced into the integral.

We neglect higher-order terms in the expansion (\ref{Aeff}), which
is justified at low densities, 
replace 
$\varepsilon_{\pm Q+k}$ that
enters $U^{(2)}$ by its leading small-$k$ term $k^{2}/(2m)$,
and  neglect the momentum and energy dependence of $U^{(4)}$, which is
appropriate in the limit of small $\Lambda$.
Then the effective low-energy action (\ref{Aeff}) 
takes the form of the Gross-Pitaevskii functional for a two-component field:
\begin{eqnarray} 
\label{AGP}
\mathcal{A}_{\Phi}&=&\int d\tau  \int dx \Big\{ \Phi^{\alpha *}(\partial_{\tau}-\mu)\Phi^{\alpha}\nonumber\\
&+&\frac{1}{2m}|\partial_{x}\Phi^{\alpha}|^{2} +
\frac{1}{2}\Phi^{\alpha *}\Phi^{\beta *} \Gamma_{\alpha\beta}^{\Lambda} \Phi^{\alpha}\Phi^{\beta} \Big\},  
\end{eqnarray}
where 
the renormalized running couplings are given by
$\Gamma_{\alpha\beta}^{\Lambda}=U^{(4),\alpha\beta}_{0,0,0,0}$. 
As shown in the main text by solving the
BS equation, in the one-dimensional case the dependence of those couplings on
the cutoff has the form
\begin{equation} 
\label{Gamma-full} 
\Gamma_{\alpha\beta}^{\Lambda}= \pi\Lambda/m
+C_{\alpha\beta}\Lambda^{2} +\ldots, \quad \Lambda\to 0.
\end{equation}

Now, consider the system of two
bosonic species $\varphi_{1,2}$  interacting via contact potential,
described by the action
\begin{eqnarray} 
\label{LiebLiniger} 
\mathcal{A}_{\rm \varphi}&=&\int d\tau \int dx 
\Big\{ \varphi^{*}_{\alpha}(\partial_{\tau}- \frac{\partial^{2}_{x}}{2m}-\mu)\varphi_{\alpha}\nonumber\\
&+&\frac{1}{2}g_{\alpha\beta}
 n_{\alpha}(x) n_{\beta}(x)\Big\},
\end{eqnarray}
where $n_{\alpha}=|\varphi_{\alpha}|^{2}$, $\alpha=1,2$ are the
densities of the two components. This is nothing but the two-component
Lieb-Liniger model. The low-energy effective theory for this model can
be again cast in the form (\ref{AGP}),\cite{Kolezhuk10} with the renormalized
couplings given by
\begin{equation} 
\label{LL-Gamma} 
\Gamma_{\alpha\beta}^{(LL)}=\frac{g_{\alpha\beta}}{1+\frac{m}{\pi\Lambda}
  g_{\alpha\beta}}=\frac{\pi\Lambda}{m}-\frac{1}{g_{\alpha\beta}}\left(\frac{\pi\Lambda}{m}\right)^{2}+\ldots,
\quad \Lambda\to 0.
\end{equation}
Comparing (\ref{LL-Gamma}) and (\ref{Gamma-full}), one can establish a
correspondence between the original lattice problem (\ref{hamB}) and
the simplified continuum model (\ref{LiebLiniger}) of the Lieb-Liniger
type, that is characterized by bare couplings $g_{\alpha\beta}$.


\section{\boldmath Symmetrized equations for $\Gamma_{11}$, $\Gamma_{12}$}
\label{app:BS-sym}

To simplify the equation (\ref{BS-gen}), 
we first symmetrize the kernel. Let us introduce the following
functions that are even in the transferred momentum $q$:
\begin{eqnarray} 
\label{sym-Gamma}
&& \Gamma_{q}^{(11)}(E)\equiv \Gamma_{q}(Q,Q;E),  \\
&& \Gamma_{q}^{(12)}(E)\equiv \Gamma_{Q+q}(-Q,Q;E)+\Gamma_{Q-q}(-Q,Q;E),\nonumber
\end{eqnarray}
then $\Gamma_{11}(E)=\Gamma^{(11)}_{q=0}$ and $\Gamma_{12}(E)=\Gamma^{(12)}_{q=Q}$.
From (\ref{BS-gen}), using the identities
$\varepsilon_{k}=\varepsilon_{-k}$, $V_{q}(k,k')=V_{-q}(k',k)$, and
$\Gamma_{q}(k,k';E)=\Gamma_{-q}(k',k;E)$, 
one readily obtains the following equation for
$\Gamma^{(11)}_{q}$:
\begin{eqnarray} 
\label{sym-G11} 
&&\Gamma_{q}^{(11)}(E)=V_{11}(Q,0)
-\frac{1}{L}\sum_{p}\frac{V_{11}(q,p)
  \Gamma^{(11)}_{p}(E)}{\varepsilon_{Q+p}+\varepsilon_{Q-p}-E},\nonumber\\
&& V_{11}(q,p)=\frac{1}{2}\big[ V_{q-p}(Q+p,Q-p)+(p\mapsto -p)\big], 
\end{eqnarray}
and a similar equation for $\Gamma^{(12)}_{q}$:
\begin{eqnarray} 
\label{sym-G12} 
&&\Gamma_{q}^{(12)}(E)=2V_{12}(q,Q)
-\frac{1}{L}\sum_{p}\frac{V_{12}(q,p)
  \Gamma^{(12)}_{p}(E)}{2\varepsilon_{p}+E},\nonumber\\
&& V_{12}(q,p)=\frac{1}{2}\big[ V_{q-p}(p,-p)+V_{q+p}(-p,+p)\big]. 
\end{eqnarray}
Explicitly, the  symmetrized kernels for our model are
\begin{eqnarray} 
\label{V11-12} 
&& V_{11}(q,p)=\begin{cases} 
V^{z}(q,p)
+U_{1}\cos p+U_{2}\cos 2p, & S\geq 1 \\ 
V^{z}(q,p) +U, \quad U\to \infty & S=\frac{1}{2} \end{cases},\nonumber\\
&& V_{12}(q,p)=\begin{cases} V^{z}(q,p) -J(p), & S\geq 1 \\ 
V^{z}(q,p) +U, \quad U\to \infty & S=\frac{1}{2} \end{cases},
\end{eqnarray}
where
\begin{eqnarray} 
\label{Vz} 
&& U_{1}=-2J_{1}\cos Q,\quad U_{2}=-2J_{2}\cos 2Q,\\
&& V^{z}(q,p)=2J_{1}^{z}\cos (q)\cos (p) +2J_{2}^{z}\cos (2q)\cos (2p) \nonumber
\end{eqnarray}

Solutions of Eqs.\ (\ref{sym-G11}), (\ref{sym-G12}) can be
now sought in the form containing only even Fourier harmonics:
\begin{eqnarray} 
\label{ansatz} 
&&\Gamma^{(11)}_{q}=A_{1}+A_{2}\cos q+ A_{3}\cos 2q,\nonumber\\
&&\Gamma^{(12)}_{q}=B_{1}+B_{2}\cos q+ B_{3}\cos 2q.
\end{eqnarray}
\begin{widetext}
This ansatz transforms the integral equations (\ref{sym-G11}), (\ref{sym-G12})
into  $3\times3$ linear systems:
\begin{eqnarray} 
 \label{L11}  
&& \begin{pmatrix}
\tau_{11}^{A}+\frac{2S-1}{2S(U_{1}+U_{2})-E} & \tau_{12}^{A} & \tau_{13}^{A}
\\
\tau_{12}^{A} & \tau_{22}^{A}+\frac{1}{2J_{1}^{z}} & \tau_{23}^{A} \\
\tau_{13}^{A} & \tau_{23}^{A} &\tau_{33}^{A}+\frac{1}{2J_{2}^{z}}
\end{pmatrix} 
\begin{pmatrix} A_{1} \\ A_{2} \\ A_{3}\end{pmatrix} = \begin{pmatrix}
  1/c_{S} \\ 1 \\ 1\end{pmatrix} \\
 \label{L12}
&& \begin{pmatrix}
\tau_{11}^{B}+\frac{2S-1}{2S(U_{1}+U_{2})-E} & \tau_{12}^{B} & \tau_{13}^{B}
\\
\tau_{12}^{B} & \tau_{22}^{B}+\frac{1}{2J_{1}^{z}} & \tau_{23}^{B} \\
\tau_{13}^{B} & \tau_{23}^{B} &\tau_{33}^{B}+\frac{1}{2J_{2}^{z}}
\end{pmatrix} 
\begin{pmatrix} B_{1} \\ B_{2} \\ B_{3}\end{pmatrix} = \begin{pmatrix}
  2/c_{S} \\ 2\cos Q \\ 2\cos 2Q\end{pmatrix} ,
\end{eqnarray}
where $c_{S}=1-\frac{E}{2S(U_{1}+U_{2})}$ for $S\geq 1$ and $c_{S}=1$
for $S=\frac{1}{2}$, and the matrix coefficients are given by
\begin{equation} 
\label{tau} 
\tau_{ij}^{A}=\frac{1}{L}\sum_{k}\frac{f_{i}f_{j}}{\varepsilon_{Q+k}+\varepsilon_{Q-k}-E},
\quad
\tau_{ij}^{B}=\frac{1}{L}\sum_{k}\frac{f_{i}f_{j}}{2\varepsilon_{k}-E}, \quad
 f_{1}=1,\quad f_{2}=\cos k,\quad f_{3}=\cos 2k\; .
\end{equation}
\end{widetext}
The solutions of those equations can be easily obtained
in a closed though somewhat cumbersome form.

\section{The continuum limit for a one-component lattice Bose gas:
  bare and renormalized interaction}
\label{app:1-comp}

Consider Bose gas on a lattice, with the single-particle kinetic energy $\varepsilon_{k}$ that behaves quadratically at small
momenta, $\varepsilon_{k}\simeq k^{2}/2m$ at $k\to 0$. Let us assume
there is an on-site interaction of strength $g_{0}$. In the
limit of vanishing density, one can write down the Bethe-Salpeter equation for the renormalized 
two-body interaction $\Gamma_{q}(k=0,k'=0;E=0)\equiv\Gamma$ in the same way as we have
done in Sect.\ \ref{sec:efftheory}, but now it is considerably
simplified since the bare interaction vertex $V_{q}(k,k)=g_{0}$
is independent of momenta, and the renormalized interaction
$\Gamma$ is independent of the transferred momentum $q$:
\begin{equation} 
\label{BS-1comp} 
\Gamma=g_{0}(1 -\Gamma I),\quad
I=\frac{1}{\pi}\int_{\Lambda}^{\Lambda_{\rm UV}} \frac{dp}{2\varepsilon_{p}}.
\end{equation}
Here $\Lambda_{\rm UV}=\pi$ is the
natural ultraviolet lattice cutoff, and $\Lambda$ is the running infrared 
cutoff. The flow eventually gets stopped at $\Lambda=\Lambda_{*}$
given in Eq.\ (\ref{Lambda*}). At low gas density $n$ the regulator
$\Lambda_{*}=\pi n/2 \ll 1$, so we are interested in the behavior of
$\Gamma(\Lambda)$ at $\Lambda\to 0$. 

For the continuum Bose gas model with contact interaction of the form
$U(x-x')=g_{c}\delta(x-x')$ (the Lieb-Liniger model), the equation for
the renormalized interaction $\Gamma_{c}$ is very similar,
\begin{equation} 
\label{BS-1comp-cont} 
\Gamma_{c}=g_{c}(1 -\Gamma_{c} I_{c}),\quad
I_{c}=\frac{m}{\pi}\int_{\Lambda}^{\infty} \frac{dp}{p^{2}}=\frac{m}{\pi\Lambda}.
\end{equation}
The running coupling in the continuum model
is thus given by
\begin{equation} 
\label{Gamma-Lieb-Liniger} 
\Gamma_{c}=\frac{g_{c}}{1+\frac{m}{\pi\Lambda} g_{c}}.
\end{equation}

Let us come back to the lattice expression (\ref{BS-1comp}).
Since $\varepsilon_{k}\to
k^{2}/2m$ at $k\to 0$, it is easy to see that the integral
$I$ in (\ref{BS-1comp}) behaves as
\begin{equation} 
\label{I-Lambda}
I\to \frac{m}{\pi \Lambda}+ I_{\rm lat} , \quad \Lambda\to 0, 
\end{equation}
where $I_{\rm lat}$ is some parameter of the order of unity that incorporates all effects of
the lattice (ultraviolet cutoff as well as the deviation of the
dispersion $\varepsilon_{p}$ from the quadratic law). The important
point is that this constant $I_{\rm lat}$, depending on the lattice
details, can be \emph{negative}: for instance, if one takes the free
particle dispersion $\varepsilon_{p}\simeq p^{2}/(2m)$ and simply
introduces the lattice cutoff $\Lambda_{\rm UV}$, then $I_{\rm
  lat}=-m/(\pi\Lambda_{\rm UV})$. 
Curiously, this constant $I_{\rm lat}$ does not appear if the particle
dispersion is a pure cosine (i.e., only nearest-neighbor hopping is
taken into account): it is easy to show that for
$\varepsilon_{p}=\frac{1}{m}(1-\cos p)$ the integral $I$ in
(\ref{BS-1comp}) is
\begin{equation} 
\label{disp-cosine}
 I=\frac{m}{2\pi\tan(\Lambda/2))}\to \frac{m}{\pi \Lambda}(1-
 \frac{\Lambda^{2}}{12} +\ldots) \quad \text{at $\Lambda\to 0$},
\end{equation}
i.e., in this case $I_{\rm lat}\to 0$.

The resulting behavior of the running coupling in the lattice problem
can be represented in the form identical to that of the continuum
model:
\begin{equation} 
\label{Gamma-lattice} 
\Gamma=\frac{\tilde{g}_{0}}{1+\frac{m}{\pi\Lambda}
  \tilde{g}_{0}},\quad \tilde{g}_{0}=\frac{g_{0}}{1+g_{0}I_{\rm lat}}.
\end{equation}
Comparing equations (\ref{Gamma-Lieb-Liniger}) and
(\ref{Gamma-lattice}), one can see that $\tilde{g}_{0}$ plays the role
of the effective bare coupling $g_{c}$ if the  original lattice problem is mapped
to the continuum Lieb-Liniger model. The value of $\tilde{g}_{0}$
will, e.g., determine the phase shifts in two-body scattering.
As long as the coupling constant
$g_{0}$ is small, such a mapping presents no problem.
However, it is easy to see that
if $I_{\rm lat}$ is negative and $g_{0}|I_{\rm lat}|$ becomes close or
exceeds $1$,
then even the sign of $\tilde{g}_{0}$ can differ from that of the lattice bare coupling
$g_{0}$. The parameter $\tilde{g}_{0}$ can be directly interpreted as
the effective
continuum bare coupling only if the condition 
\begin{equation} 
\label{appl-cont-1comp}
|\tilde{g}_{0}|m \ll 1 
\end{equation}
is satisfied, otherwise one has to look at the dressed coupling $\Gamma(\Lambda_{*})$.


\section{Schr\"odinger's equation for the two-magnon scattering problem}
\label{app:twomagnon}

We set the magnetic field at the saturation value,\cite{commentSaturationField}
$H=H_S=2S(J_1^z+J_2^z)+S(J_1^2+8J_2^2)/4J_2$ and consider a general
two-magnon state that can be written in the following form (here we use a
normalization different from that of
Ref.\ \onlinecite{Arlego+11}):
\begin{equation}
 |\Psi_{2M}\rangle =\sum_{n,m\ge n}c_{n,m} 
\frac{S^-_n S^-_m|F\rangle }{\langle F| S^{\dagger}_m S^{\dagger}_n \,\,   S^-_n S^-_m|F\rangle} ,
\end{equation}
where $|F\rangle$ is the fully polarized state. The total
quasi-momentum $K$ is conserved, therefore one can separate the center
of mass motion by setting
$c_{n,m}=\exp(i\frac{n+m}{2}K)C_{m-n}$. Solving the Schr\"odinger
equation $\mathcal{H} |\Psi_{2M}\rangle =E |\Psi_{2M}\rangle$ for
scattering states in the thermodynamic limit, one can identify the
energy $E=\varepsilon(k_{1})+\varepsilon(k_{2})+E_F$, where
$E_F=N(S^2(J_1^z+J_2^z) -H_SS)$ is the energy of the fully polarized
state, as
\begin{eqnarray} 
\label{Escat}
E&=&\Omega_{0}-2SJ(Q)+E_F,\nonumber\\
\Omega_{0}/4S &=& J_1\cos(k)\cos\frac{K}{2} + J_2\cos(2k)\cos(K),
\end{eqnarray}
where $k_{1}$, $k_{2}$ are the magnon momenta,
$k=\frac{1}{2}(k_{1}-k_{2})$ is one half of the relative momentum, and
$K=k_{1}+k_{2}$ is the total momentum. The
Schr\"odinger equation leads to the following linear system for the
amplitudes $C_{r}$:
\begin{widetext}
\begin{eqnarray}
\label{Sch}
\Omega_0C_0&=&2\sqrt{S(2S-1)}(J_1C_1 \cos{\frac{K}{2}}+J_2C_2 \cos{{K}} ),\nonumber\\
(\Omega_0-J^z_1)C_1&=&2J_1\sqrt{S(2S-1)}C_0\cos{\frac{K}{2}}+2SJ_1C_2 \cos{\frac{K}{2}}+2SJ_2(C_1+C_3) \cos{{K}},\nonumber\\
(\Omega_0-J^z_2)C_2&=& 2J_2\sqrt{S(2S-1)}C_0\cos{{K}}+2SJ_2C_4\cos{{K}}+2SJ_1(C_1+C_3)\cos{\frac{K}{2}},\nonumber\\
\Omega_{0}C_{r} & = &2SJ_1\cos{\frac{K}{2}} \left(C_{r+1}+C_{r-1}\right)+  2SJ_2\cos{K}\left(C_{r+2}+C_{r-2}\right)\quad(r\geq3) .
\end{eqnarray}
\end{widetext}


\section{Scattering phase shift in the interspecies scattering problem}
\label{app:interscattering}

To extract the phase shift for the interspecies scattering problem
from the solution (\ref{SStateDegenerate1}), we put
$k=Q+p$ and consider the limit of small deviation
from the minimum of dispersion $p\to 0$, then $\tilde k=Q-p+O(p^2)$ and the scattering state
Eq.(\ref{SStateDegenerate1}) can be rewritten as follows,
\begin{eqnarray}
\label{interscatphase}
C_{r} &&= \sqrt{\cos^2{\alpha}+(v+\sin{\alpha})^2}\cos (Qr)\cos (pr+\delta_{12})\nonumber\\
&&-\sqrt{\cos^2{\alpha}+(v-\sin{\alpha})^2}\sin{(Qr)}     \sin(pr-\gamma) 
\end{eqnarray}
where 
\begin{eqnarray}
\label{inter-delta}
\delta_{12}&=& -\arccos{\frac{\cos{ \alpha }}{  \sqrt{\cos^2{\alpha}+(v+\sin{\alpha})^2}  }},\nonumber\\
\gamma&=& -\arccos{\frac{\cos{ \alpha }}{  \sqrt{\cos^2{\alpha}+(v-\sin{\alpha})^2}  }}.
\end{eqnarray}
From the form (\ref{interscatphase}) one can easily guess that for
$J_1=0$ ($Q=\pi/2$) the scattering phase shift is $\delta_{12}$. This
is indeed so, because to obtain the phase shift we have to put $r=2n$ and the
second term in Eq.\ (\ref{interscatphase}) vanishes (for odd values
$r=2n+1$ there is no scattering involved since at $J_{1}=0$ the system
corresponds to two decoupled chains).  

For $Q=0$ (i.e., at the
ferromagnetic Lifshitz point $\beta=-4$) and $Q=\pi$ (the
antiferromagnetic Lifshitz point $\beta=4$) it is also easy to
recognize $\delta_{12}$ as the scattering phase shift.

Furthermore, for general values of $\beta$ one can show that
$\lim_{p\to 0} \alpha=\pm \pi/2$ and $\lim_{p\to 0} v=
\pm 1$, so that $\lim_{p\to 0}(v-\sin{\alpha})= \lim_{p\to 0}
\cos{\alpha} = 0$. This corresponds to the fact that  scattering states vanish when $p\to0$, i.e., when the
magnon momenta tend to the dispersion minima $k_{1}\to Q$, $k_{2}\to -Q$,
and reflects the
 ferminionization of excitations
forming a single Fermi sea, as mentioned in the main text. 
It is easy to see that
 the second amplitude in Eq.\ (\ref{interscatphase}) tends to zero,
 $\lim_{p
  \to 0}\sqrt{\cos^2{\alpha}+(v-\sin{\alpha})^2} = 0$, while the first
amplitude tends to a constant,
$\lim_{p \to 0}\sqrt{\cos^2{\alpha}+(v+\sin{\alpha})^2} =2$. 
 Thus, the scattering phase shift for interspecies
scattering at $p \to 0$ is given by $\delta_{12}$ for any $\beta$.

One can also show that 
the two following expressions are equal to each other, 
\begin{equation}
\lim_{p \to 0} \frac{\cot{(\delta_{12})}}{p}=\lim_{p\to 0} \frac{\cot{(\alpha)}}{2p}.
\end{equation}
In other words, if $\alpha\to - \pi/2+2a_{12}p$ at $p\to 0$, then
$\delta_{12}\to - \pi/2+ a_{12}p$.



\begin{thebibliography}{10.}
\bibitem{noteDM} The use of Dyson-Maleev representation which leads to
  a non-hermitean bosonic Hamiltonian is not
  important but merely convenient. The exact ladder structure of the
  Bethe-Salpeter equation means that we are in fact solving a two-body
  problem (repeated scattering of two particles). One could thus use
  the Goldhirsch representation\cite{Goldhirsch} for spin operators and neglect terms
  of higher than quartic order in the bosonic Hamiltonian, which leads
  to a hermitean Hamiltonian for general spin $S$, but with slightly
  more cumbersome coefficients. We have explicitly checked that the final results do
  not depend on the chosen representation.
\bibitem{commentSaturationField} If magnons form bound states, the
  actual saturation field (minimal strength of magnetic field needed
  to fully polarize the ground state) differs from $H_s$ by the
  magnon binding energy.
\bibitem{note-oldpaper} In Ref.\ \protect\onlinecite{KV05}, there is a
mistake in Eq.~(21), leading to the erroneous impression that in the
isotropic chain $g_{12}$
may become smaller than $g_{11}$ in the semiclassical limit $S\to\infty$.
\bibitem{note-1sea} The property of refermionization with a common
  Fermi sea in the TLL2 phase applies only in the limit of vanishing
  density, when it is not possible to use the Luttinger
  liquid description (in the dilute limit the fixed point of the
  renormalization group flow is a theory of non-relativistic free
  fermions with the dynamical critical exponent $z=2$). This has to be
  contrasted with the situation at finite density, when the
  hydrodynamic regime develops and the TLL2 state is governed by the
  fixed point of two decoupled Gaussian models (two copies of Tomonaga-Luttinger
  liquids).
\bibitem{note-s12fm} Actually, the lowest energy of a two-magnon bound
  state in the isotropic ferromagnetic $S=\frac{1}{2}$
  frustrated chain is reached at $K=\pi-\delta(\beta)$ with $\delta \ll 1$, for $-4 <\beta< -2.67$,
  and for $-2.67<\beta< 0$ the minimum of the bound state dispersion
  lies at $K=\pi$.\protect\cite{KuzianDrechsler07} For that reason,
  our theory is not directly applicable to this case.
\end{thebibliography}

\end{document}